\documentclass[]{article}
\usepackage[utf8]{inputenc}
\usepackage{amsmath}
\usepackage{blkarray}
\usepackage{caption}
\usepackage{authblk}
\usepackage{graphicx}
\usepackage{float}
\usepackage{url}
\usepackage[margin=0.5in]{geometry}

\title{\Huge An agent-based modelling framework to study growth mechanisms in \textit{EGFR-L858R }mutant alveolar type II cells

\Large Short title: Early effects of oncogenic mutation on lung stem cell behaviour}

\author[1, \dag]{Helena Coggan}
\author[2, \ddag]{Clare E. Weeden}
\author[3, \dag, \S]{Philip Pearce}
\author[4, \dag, \S]{Mohit P. Dalwadi}
\author[5, \ddag]{Alastair Magness}
\author[6, \ddag, \ss]{Charles Swanton}
\author[7, \dag]{Karen M. Page}
\affil[1]{ORCID 0000-0002-7798-9318}
\affil[2]{ORCID 0000-0002-1561-1416}
\affil[3]{ORCID 0000-0001-5788-3826}
\affil[4]{ORCID 0000-0001-5017-2116}
\affil[5]{ORCID 0000-0001-9876-3863}
\affil[6]{ORCID 0000-0002-4299-3018}
\affil[7]{ORCID 0000-0003-4189-4664}
\affil[ \dag]{University College London, Department of Mathematics}
\affil[ \ddag]{Cancer Evolution and Genome Instability Laboratory, The Francis Crick Institute}
\affil[ \S]{UCL Institute for the Physics of Living Systems}
\affil[ \ss]{Cancer Research UK Lung Cancer Centre of Excellence, University College London Cancer Institute, and Department of Oncology, University College London Hospital}

\begin{document}

\maketitle

\section{Abstract}

Mutations in the epidermal growth factor receptor (EGFR) are common in non-small cell lung cancer (NSCLC), particularly in never-smoker patients. However, these mutations are not always carcinogenic, and have recently been reported in histologically normal lung tissue from patients with and without lung cancer. To investigate the outcome of EGFR mutation in healthy lung stem cells, we grew murine alveolar type-II organoids monoclonally in a 3D Matrigel. Our experiments showed that the \textit{EGFR-L858R} mutation induced a change in organoid structure: mutated organoids displayed more `budding', in comparison to non-mutant controls, which were nearly spherical. We perform on-lattice computational simulations, which suggest that this can be explained by the concentration of division amongst a small number of cells on the surface of the organoid, which may arise from several possible biological mechanisms. These results suggest that the L858R mutation produces structures which expand quickly from surface protrusions. We are currently unable to distinguish the cell-based mechanisms that lead to this spatial heterogeneity in growth, but suggest a number of future experiments which could be used to do so. We suggest that the likelihood of L858R-fuelled tumorigenesis is affected not just by random fluctuations in cell fitness, but by whether the mutation arises in a spatial environment that allows mutant cells to reproduce without being forced to encounter each other. These data may have implications for cancer prevention strategies and for understanding NSCLC progression.

\section{Author Summary}
Cancer is driven by the development of genetic mutations. Some mutations which appear in aggressive lung cancers, particularly in people who have never smoked, have also been found to exist quite harmlessly in perfectly healthy people. Although inflammatory cytokines have been highlighted as important promoters of tumour formation, it is unclear what additional stimuli are required in order to drive a `normal cell' harbouring an oncogenic mutation into an invasive tumour. To examine this, we looked at the behaviour of stem cells with an activating mutation in EGFR, L858R, when they were given all the nutrients and space required to grow uninhibited in three dimensions. We used computational simulations to model their growth, and predicted that these organoids seem to concentrate division amongst a small number of surface cells. We hypothesise that in the very early stages of cancer development, this mutation allows cells to quickly form invasive protrusions and suppress the division of neighbouring cells. This also suggests that the success of these pre-cancerous cells depends on their spatial environment and surrounding cell ecology. We hope that this insight into early cancer development will drive more research into the consequences of cell-cell interaction dysfunction on early tumour initiation.
\section{Introduction}

Lung cancer is the leading cause of cancer death worldwide \cite{Sung2021}. Approximately a quarter of lung cancer patients are `never-smokers' (classified as those who have smoked fewer than 100 cigarettes in their lifetime) \cite{Sun2007}, and this proportion appears to be increasing \cite{Pelosof2017}. Never-smoking lung cancer is also more common in women than men \cite{Schabath2019}.

Lung adenocarcinomas are thought to arise from alveolar type II (AT2) cells within the healthy alveolar epithelium \cite{Sutherland2014}. The causes of lung cancer in never-smokers are unclear, although previous studies have highlighted germline genetics \cite{Zhang2021} and exposure to external factors such as infections and radon \cite{Corrales2020}, as well as ambient air pollution \cite{Hill2023}. Some specific lung cancer-associated mutations, in particular in EGFR, are known to occur more frequently amongst patients with no history of smoking \cite{Chapman2016}, and are found as clonal driver mutations in lung adenocarcinomas \cite{Jamal2017}. However, their presence alone is insufficient for tumorigenesis: Swanton and colleagues have recently reported that 18\% of normal lung tissue samples in patients both with and without lung cancer were found to carry EGFR mutations \cite{Hill2023}. In order to determine the conditions necessary for lung cancer initiation, we must look further, to the cellular and environmental context in which carcinogenic mutations arise.

It is well-known that the emergence of cancer is probabilistic, and that tumours can take a variety of winding evolutionary paths to acquiring their six characteristic hallmarks \cite{Hanahan2000}. No mutation is guaranteed to cause cancer. The number of cells in the human body is such that every possible genetic mutation is likely to exist in at least one cell in everyone \cite{Traulsen2010}; the fact that some people do not develop cancer suggests that not all of these mutations lead to disease in every case. The reproductive fitness of a mutant cell depends not just on tissue context \cite{Haigis2019}, but also on spatial constraints \cite{West2021} and the ecology of the wider cell population \cite{Gatenby2014}. Subclonal cells interact with each other and compete for resources, giving rise to highly genotypically heterogeneous tumours \cite{Tabassum2015}, \cite{Gerlinger2012}. Within mathematics, the field of evolutionary game theory describes the effect of such interactions on population composition, and has seen a variety of recent applications to cancer modelling \cite{Coggan2022}. But to make full use of these theories we need strategic quantification: in order to predict a phenotype's success, we must determine how it interacts with other phenotypes, and thus how its population will rise and fall within a tumour population. To identify the evolutionary benefit of a particular mutation arising in a specific cell type, then, we must consider the effect on that cell's behaviour. Only then will we be able to predict its contribution to tumorigenesis.

The aim of this study is to analyse the phenotypic changes conferred by the \textit{EGFR-L858R }mutation on alveolar type-II cells. The single nucleotide substitution L858R in exon 21 comprises around 40\% of all EGFR mutations in lung cancer patients \cite{Shigematsu2006}. It is a missense mutation, affecting the intracellular kinase domain of the epidermal growth factor receptor (EGFR), which in turn affects a cell's response to many common growth factors. Biochemical and structural studies have shown that the \textit{EGFR-L858R} mutation promotes dimerisation \cite{Shan2012} and destabilises the inactive formation \cite{Yun2007} of EGFR, and thus causes abnormal levels of receptor activity. The function of this mutation is also highly clinically relevant. \textit{EGFR-L858R} mutations confer sensitivity to treatment with EGFR tyrosine kinase inhibitors, although acquired resistance can rapidly develop \cite{Hong2019}, \cite{Reita2021}. In addition, L858R has been shown to enhance invasiveness in adenocarcinomas \cite{Tsai2015}, and mice with L858R mutations induced in alveolar type-II cells rapidly develop diffuse carcinomas \cite{Politi2006}. It is this particular aspect of mutant behaviour that we address in the present study.

We infer an invasive mutant phenotype using an agent-based model (ABM) of organoid growth in a three-dimensional organoid culture, by comparing the structures of spheroids grown monoclonally from mutant and wild-type AT2 cells. ABMs have been common throughout systems biology for many years \cite{An2009}, and are useful for describing systems where individual `agents' (here AT2 cells) interact and reproduce probabilistically according to defined rules. Our focus is on the emergence of a characteristic `budding' structure in some mutant cell clusters, as opposed to their wild-type counterparts, which grow spherically. These organoids are composed of thousands or tens of thousands of cells, and are smaller than is required for the development of vascularity. The field of mathematical modelling of avascular tumour growth is larger than can be fully summarised here (see \cite{Roose2007, Byrne2009, Byrne2012} for an introduction). One common methodology is a `continuum' approach, where cell density is described as a continuous scalar field, which evolves with time. Cell-cell interactions and surface-dominated growth can be included in a continuum model \cite{Breward2002, Ciarletta2013}, and the resulting approach is amenable to mathematical analysis, which has shown surface-dominated growth to drive the formation of non-spherical structures. These models are particularly advantageous in the study of reaction-diffusion systems involving nutrient gradients which evolve with time \cite{Miura2008}, and when analysing the contribution of various biomechanical processes to general morphological instabilities \cite{Giverso2016}.  While these approaches are broadly useful, in many situations the problem we are modelling is three-dimensional, extremely asymmetric, and involves fully developed and morphologically specific protrusions, which require high spatial granularity to model. Here, where we wish to test nonlinear growth laws where the discrete number of neighbours a cell has is important, and where the number of cells involved is relatively small (on the order of thousands or tens of thousands), it is both conceptually and computationally easier to model the location and action of each cell directly. ABMs (sometimes also called `cellular automata models') have been used for almost two decades to construct simulations of tumour growth in which cells are considered as discrete agents making `cell-fate decisions' in response to changing environmental conditions such as oxygen depletion \cite{Gerlee2007}. In recent years, ABMs have been used to shed light on increasingly complex systems of cell-cell interaction \cite{Wang2015, Metzcar2019, Sadhukhan2021, Sivakumar2022}. We continue that approach in this study.

Here we develop an algorithm in which space is divided into a fixed three-dimensional lattice, whose points may be either occupied or empty, to simulate cells as they divide and push each other aside. This approach significantly speeds up computation times compared to off-lattice simulation (where the distance between of each pair of cells must be recalculated at each timepoint), which allows for thorough hypothesis testing and parameter sweeps. Our study is driven by the principles of mathematical modelling, wherein complex biological processes are abstracted into high-level mathematical functions. Here, we consider functions which take the number of neighbours a cell has and outputs the probability that that cell will divide in a given timestep. Whether or not that division actually occurs is decided stochastically, and over time this results in a simulated organoid with an observable morphology. We can alter the forms and parameters of these functions to represent different hypotheses concerning cell growth. The values of these parameters may not correspond to specific biological quantities, but by altering these values-- how many neighbours a cell must have before its probability of division begins to decline, for example-- we can change the hypothetical rules for cell growth which apply to our system. We can then determine which set of models and parameter regions results in the correct morphologies, and infer the validity of the corresponding rules, without needing to know the biological mechanisms by which they are enforced.

We test various growth hypotheses, including those where cell division is limited by differentiation and by the presence of surrounding cells, and compare the simulated organoids to experimentally observed morphologies. We find that the structural difference between mutant and non-mutant organoids can be explained by one of four possible models, all of which result in the concentration of growth amongst a small number of surface cells. In the first such model, this `surface-localisation' arises from very high levels of differentiation amongst all cells combined with more efficient use of available nutrients by mutant cells. In the second, `budding structures' arise from long-range inhibition of cell division by their neighbours, combined with an increase in division rate by mutant cells. In the third and fourth, the mutation induces short-range inhibition of neighbouring cells through mechanotransduction, allowing the formation of surface protrusions without any assumed increase in cell fitness. In the fourth, this induced inhibition is long-range, through either nutrient depletion or the active secretion of some inhibitor. In all hypotheses, cells do not have to be completely surrounded to suffer inhibition, and are stopped from dividing even when half of their surface is in contact with the nutrient-rich Matrigel. The overall effect is to suppress division in all cells except those on the surface, localising division within invasive protrusions. We hope this will shed light on the origin of invasive \textit{EGFR-L858R-}positive cancers, and the mechanisms by which they emerge from healthy tissue.

\section{Results}

\subsection{Experimental design}

To obtain data for comparison with our hypothesis-testing simulations, we used a genetically engineered mouse model of EGFR-L858R-driven lung adenocarcinoma, as described in \cite{Hill2023} (Rosa26\textsuperscript{LSL-tTa/LSL-tdtomato}; \textit{TetO-huEGFR}-\textsuperscript{L858R} mice; referred to as ET mice). Here, both the \textit{EGFR-L858R} transgene and tdTomato fluorescent reporter protein are only expressed upon delivery of Cre recombinase. Lox-stop-lox tdTomato mice were used as control (\textit{Rosa26}\textsuperscript{LSL-tdTomato/LSL-tdTomato}; referred to as T mice). Alveolar type II cells were purified from non-Cre treated T and ET mice, before inducing expression of the oncogene and/or tdTomato in vitro with adenoviral-CMV-Cre incubation using published methods \cite{Major2020},  \cite{Dost2020}. 10,000 AT2 cells were seeded in each organoid assay along with 50,000 supporting lung fibroblasts \cite{Choi2020}; in each experiment only some AT2 cells grew into organoids (with organoid forming efficiency of around 1-2\%). After 14 days, organoids were extracted from matrigel, stained with antibodies and analysed by 3D wholemount confocal microscopy \cite{Dekkers2019}. These analyses revealed a characteristic `budding' structure occurring in mutant organoids, but not wild-type organoids (Figure 1). We quantify this effect by analysing the \textit{circularity} of each organoid. An organoid with cross-sectional area \(a\) and perimeter \(p\) has a circularity \(C= \frac{4 \pi a}{p^2}\). A perfectly circular organoid has  \(p=2\sqrt{\pi a}\) and so \(C=1\). As an organoid acquires protrusions, its perimeter increases relative to its area, and so its circularity increases. A shape with infinite perimeter and finite area (i.e. a fractal) has \(C=0\). This is a dimensionless coefficient and so is invariant to the scale at which the organoid is observed. We use the image processing libraries scikit-image \cite{scikit-learn} and Shapely \cite{shapely2007} to automatically find the edges of each organoid and calculate its perimeter and area. This analysis reveals a fairly clean delineation between mutant and non-mutant organoids, with all but one mutant organoid with a circularity  \(C<0.78\) (Figure 1). Since only mutants have a circularity below this value, we will use this as our criterion for whether a simulated organoid is `mutant-like'.

\begin{figure}[H]
  \includegraphics[width=\linewidth]{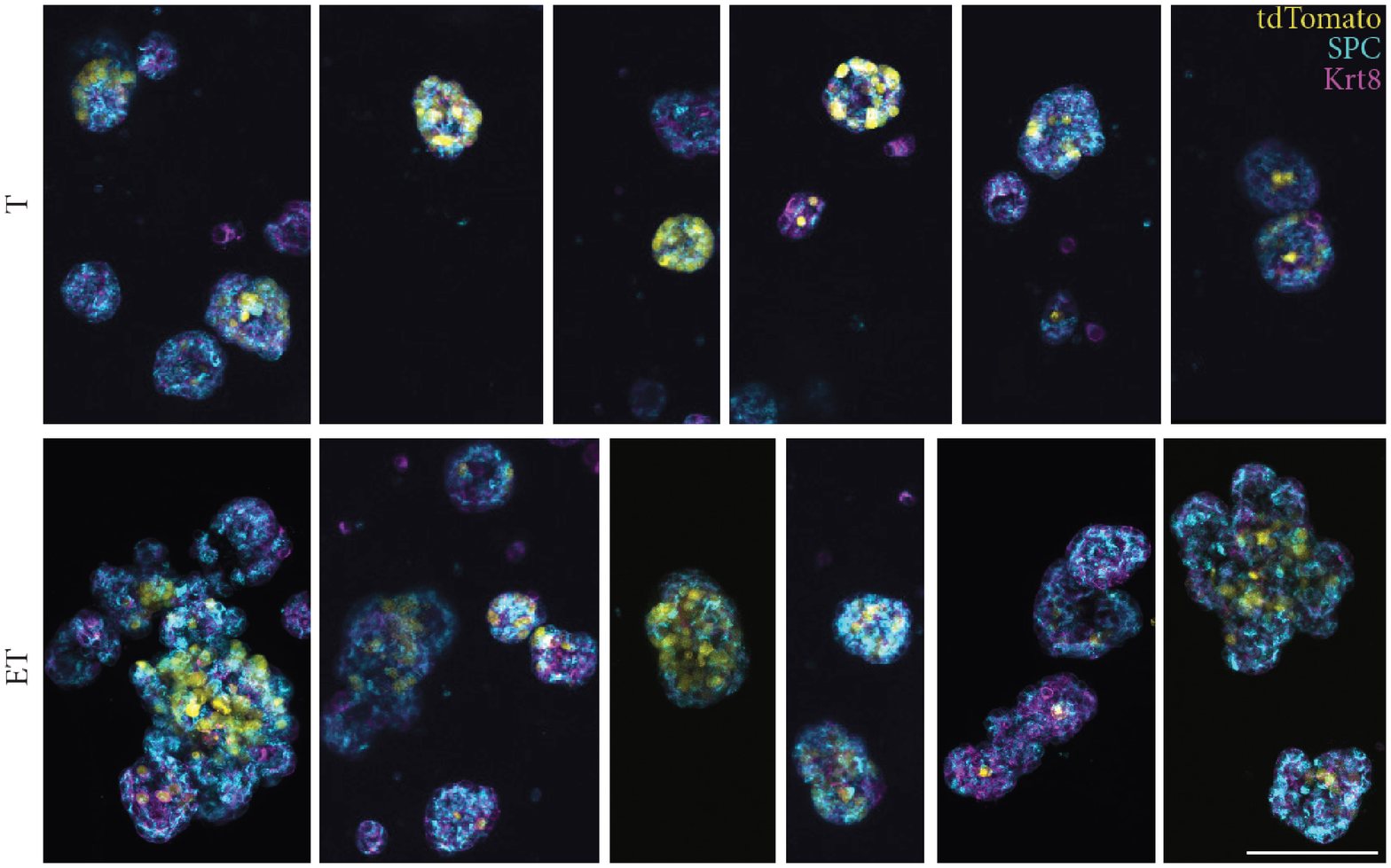}
  \includegraphics[width=\linewidth]{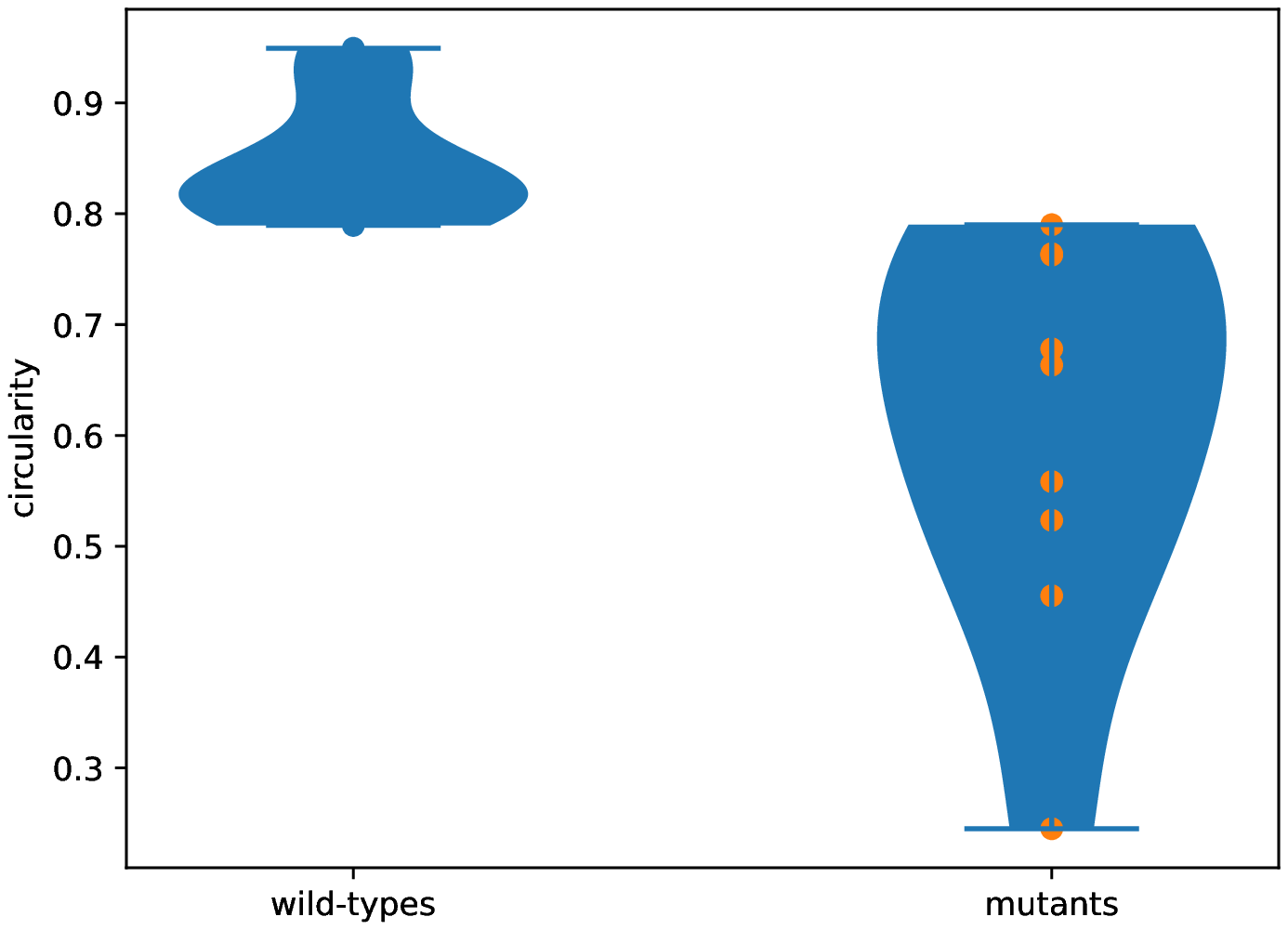}
  \caption{Top: Representative 3D confocal microscopy of AT2 organoids from T mice (top panel) and ET mice (bottom panel). Organoids stained with anti-surfactant protein C (SPC, cyan) and anti-keratin 8 (Krt8, magenta), endogenous tdTomato expression. Results shown from 1 experiment comprising 3 mice; n = 2 independent experiments. Scale bar represents 100 microns. Bottom: a violin plot of the circularity of observed organoids.}
\end{figure}

\subsection{Simulation design}
To simulate the development of these structures, we use an on-lattice algorithm. In this framework, 3D space is divided into a 30x30x30 grid (adjusted to 40x40x40 where necessary to accommodate an expansive structure). At the end of each simulation step, every lattice point is either occupied by one cell or by none. Each lattice point (except those on the grid's faces) has 26 Moore neighbours, each of which is considered equally `close' to it within the algorithm. This ensures that when cells are pushed aside by new cells in a simulation step, they are not constrained to move only in six directions (up, down, left, right, forwards or backwards) but can move in 26, allowing the simulation of much more realistic organoid structures. A cell may therefore have up to 26 occupied neighbours. Further model assumptions, and their experimental justification, are summarised in Table 1.

\begin{table}[H]
\centering
\begin{tabular}{|p{0.4\linewidth} | p{0.4\linewidth}|}
\hline
\textbf{Model assumption}                                                                                  & \textbf{Experimental justification}                                                                                                                                                                                                                                                \\ \hline
All cells have the same volume (approximately \(80 \mu m^3\))                                              & The Matrigel is soft enough that cells are not significantly compressed and may push each other aside when they divide to maintain constant volume. The figure is extrapolated from measurements of areas of individual cells, approximating cells as spherical                    \\ \hline
Cells remain in contact with their organoid of origin, and do not drift freely through the Matrigel        & Organoids are cohesive and contiguous; `clouds' of drifting fluorescent cells are not observed                                                                                                                                                                                      \\ \hline
Organoids grow independently and there is no crosstalk between them                                        & Top-down images of the well suggest that the organoids are seeded very far apart from each other relative to their diameter                                                                                                                                                        \\ \hline
External nutrient concentration does not deplete over time                                                          & Nutrients are regularly replenished by the experimenter                                                                                                                                                                                                                           \\ \hline
Each cluster arises from a single cell& Fusion of growing organoids is not experimentally witnessed                                                                                                                                                                                                                        \\ \hline
\end{tabular}

\caption{Assumptions made when modelling the system computationally, and their experimental justifications.}
\end{table}

The algorithm works as follows. At the start of the simulation only a single cell is occupied at the centre of the grid. At every timepoint, each existing cell either divides or does not, according to some probabilistic function of the number of cells in its neighbourhood (which we vary according to the hypothesis being tested). If a cell divides, that lattice point is instantaneously occupied by two cells. This is referred to as the \textit{reproduction} step. 

  The next step in the algorithm is \textit{spatial adjustment}, where cells on multi-occupied lattice points are shifted around until they find empty points to settle on, mimicking the physical process of cells pushing each other aside. This step is iterative and repeats until every cell has its own point. At each iteration of the adjustment step, one cell from every multi-occupied point is moved to one of its neighbouring points. The rules for choosing a lattice-point to move to are as follows.
\begin{enumerate}
\item If there is are any empty neighbour-points, choose between them at random with probability \(p \propto e^{\tau n_1}\), where \(n_1\)is the number of occupied lattice-points in the \textit{neighbour-point's} immediate neighbourhood, and \(\tau>0\) is a surface-tension-like parameter.
\item If there are no empty neighbour lattice-points, choose one at random with probability \(p \propto e^{-\gamma  n_0}\), where \(n_0\) is the number of cells currently occupying it and \(\gamma>0\) is a `repulsion parameter'.
\end{enumerate}
The `effective surface tension' parameter \(\tau\) controls the cohesion of cluster shapes and is designed to mimic the effect of cell-cell adhesion. When it is high, cells will preferentially move to empty points next to occupied points. This is a variable parameter and can be used to qualitatively adjust the amount of surface cohesion in an organoid. The repulsion parameter \(\gamma\) is included to dissuade cells from moving between double-occupied points. Cells are pushed through the cluster to minimise compression, and so will move preferentially to points with fewer cells occupying them. This parameter is designed to mimic the effect of a pressure gradient and is kept high (\(\gamma = 1\)). 

A key benefit of this algorithm is that whether or not two lattice-points are neighbours is only ever calculated once, when the grid is generated, and does not need to be worked out again with every simulation (or, worse, with every simulation step). The number of cells touching a focal cell can be calculated by looking up the list of its neighbouring lattice-points and checking whether each of them is occupied. It does not require calculating the distance between the focal cell and every other cell in the organoid. Therefore, its computational cost does not scale with the square of the current number of cells of the cluster, as it would in an off-lattice simulation, but linearly with the number of occupied cells in the system. This means that clusters of many tens of thousands of cells can be simulated in a few seconds, in comparison to equivalent off-lattice simulations, which struggle to simulate more than a few hundred cells efficiently. This computational efficiency allows for quick parameter sweeps.

Clusters are simulated until they reach a certain experimentally realistic physical size, comparable to the observed organoid sizes (a maximum diameter of roughly \(100 \mu m\), which generally requires around 5000 or 10000 cells, depending on the division probability function under investigation). The structure of the resulting organoid is examined and compared to experimental observations. 

We use this simulation framework to investigate two broad classes of plausible hypotheses concerning organoid growth. In the first class, we assume that the probability that a cell will divide depends only on the number of cells in its immediate neighbourhood. This hypothesis class is motivated by noting that the Matrigel is highly diffusive (eg. VEGF has a diffusion coefficient in it of the order of \(10^6 \mu^{2} m /h\) \cite{Miura2009}) and so concentration gradients in nutrients and growth factors are difficult to maintain on the timescale of cell division. Cells suppress or induce each other's division only when their surfaces touch. These mechanotransduction-based hypotheses are covered in Sections 4.2 and 4.3.

In the second class of hypotheses, we relax this assumption. We suppose instead that a `depletion field' exists around each cell, reducing the ability of nearby cells to divide. In this hypothesis class, the influence of one cell on another decreases with distance but never vanishes entirely, such that non-neighbouring cells can suppress each other's growth through either directly inhibitory signalling or competition for resources. These long-range-interaction-based hypotheses are covered in Section 4.4.

Under both hypotheses (i.e. assuming either short- or long-range interactions), mutant-like organoids are created when this mutual inhibition is strong enough to concentrate growth amongst a small number of surface cells, allowing the development of invasive protrusions. These allow mutant organoids to reach a larger maximum radius using fewer cells than non-mutant organoids, resulting in a highly efficient mode of spatial expansion and suppressing the growth of competing cells. These results are discussed in more detail below.

\subsection{Nutrient-absorption or anchorage-dependent growth alone are insufficient to produce mutant-like organoid structures}

\subsubsection{Absorption-based growth}

A plausible hypothesis for the emergence of `budding' structures is that cells on the surface of the organoid (i.e. those with fewer neighbouring cells) have better access to nutrients and so have a higher probability division. We hypothesise here that all cells have sufficient access to nutrients to maintain a non-zero division rate even when completely surrounded (see Table 1); later we will investigate a regime in which surrounded cells are modelled as having a negligible division rate.

We test a division probability which is a monotonically increasing function of the number of empty lattice-points in a cell's immediate environment, \(n\), which eventually saturates on physical grounds \cite{EUNGDAMRONG2004}. Within this framework, the probability \(p\) that a cell will divide in a given timestep \(dT\) is

\[p_{abs}(\alpha, \beta, h, c, n, dT) = \left(\alpha + \frac{\beta}{1+e^{-h(c-(26-n))}}\right)dT \]
where \(\alpha >0\) is a `baseline' division rate possessed by completely surrounded cells; \(\beta\) is the strength of the nutrient-derived division-rate boost (such that the maximum possible division rate of an isolated cell is \(\alpha + \beta\)); \(c\) is the threshold number of occupied neighbours at which the proliferative capacity of a cell decreases; and \(h\) is the `steepness' of that saturation, and 26 is the maximum number of possible neighbours. The variation of this function with \(h\) and \(c\) is illustrated in Figure 2.

\begin{figure}[]
 
  \includegraphics[width=0.49\linewidth]{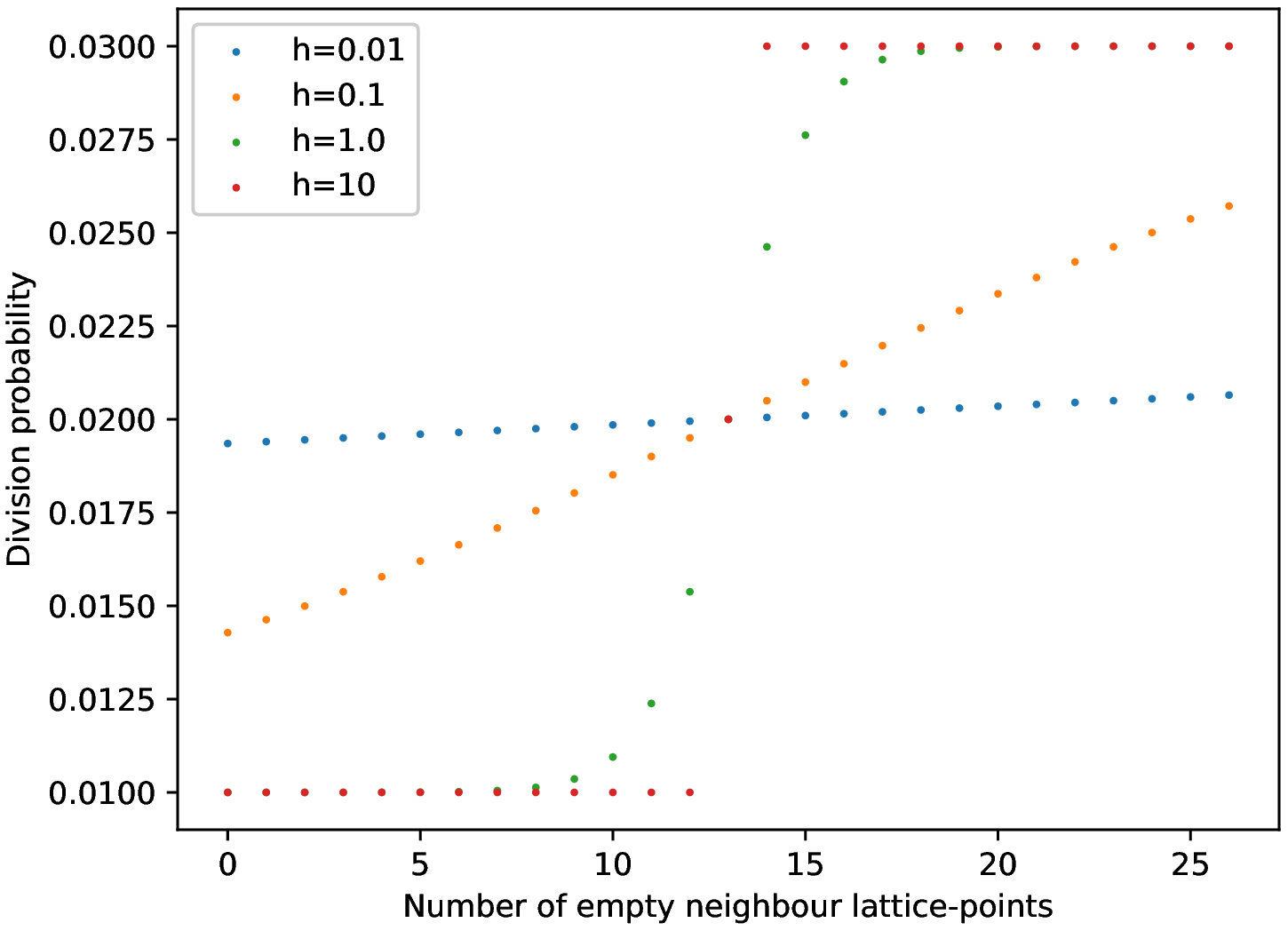}
  \includegraphics[width=0.49\linewidth]{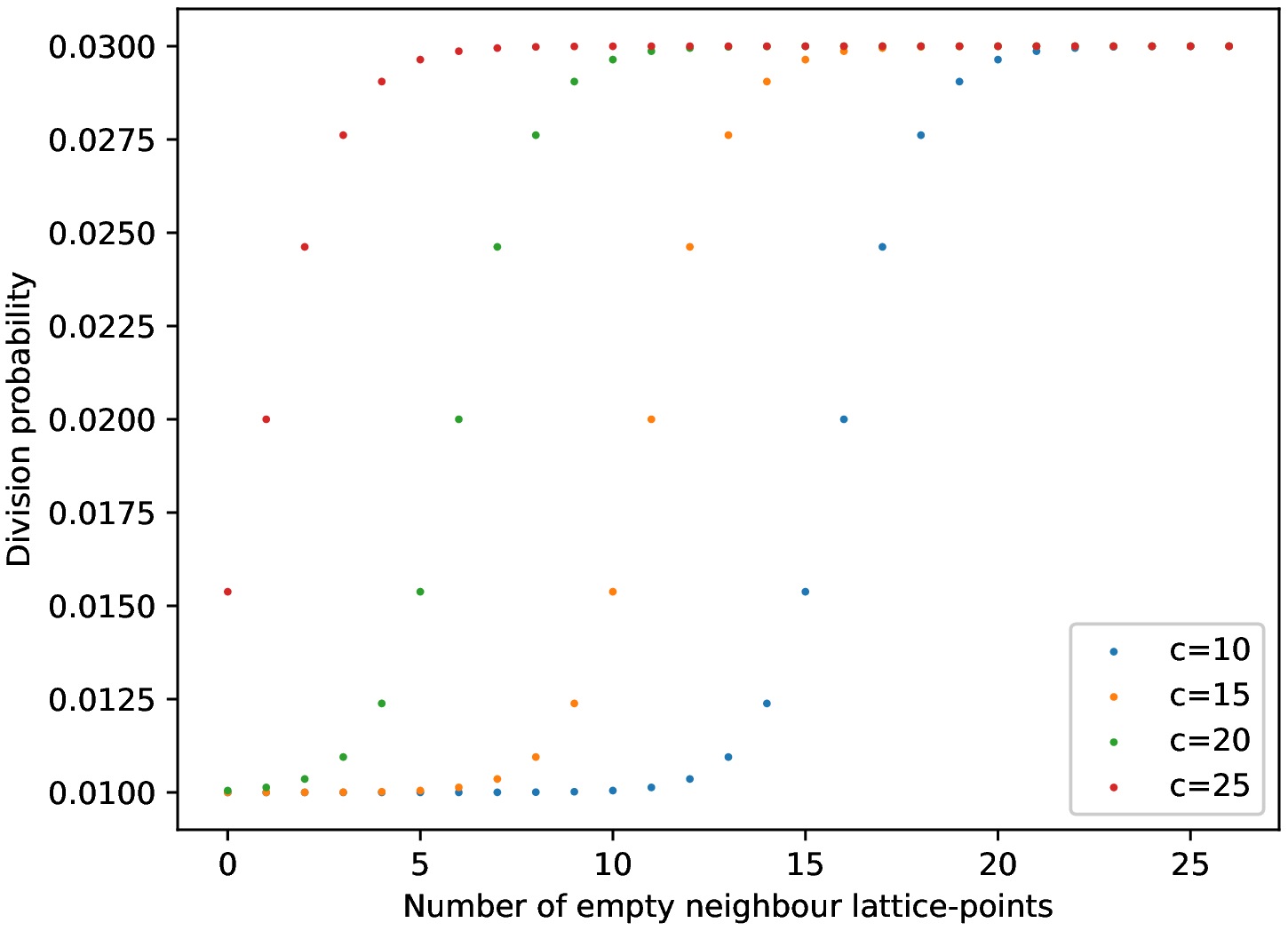}
  \caption{An illustration of the nutrient-absorption-based probability function with \(\alpha = 1, \beta = 2\), i.e. the probability of reproducing in a time interval \(dT=0.01\) (days). On the left, \(c=13\) empty neighbours and the steepness \(h\) is varied. On the right, \(h=1\) and \(c\) is varied.}
\end{figure}

Throughout these experiments \(dT = 0.01\) days, or just under 15 minutes, to ensure the number of cells generated in any given step is not too high. In Figure 3, for example, we see that even completely isolated cells only ever have a 3\% probability of reproducing per timestep. In general we consider very large values of \(h\) to be biologically implausible, as they would imply sudden changes in a cell's ability to proliferate if it acquires one or two extra neighbours. We therefore restrict ourselves to values of \(h\) of order 1 or below. In a system where \(\beta = 0\) and \(\alpha > 0\), all cells would have equal ability to divide and the resulting organoid would be necessarily spheroidal. To test whether any other sets of parameters can result in the development of secondary spheroids, we fix \(\alpha\) and vary other parameters to test the possible results of the growth mechanism. Only the values of these parameters relative to \(\alpha\) matters; increasing or decreasing all of them by some multiplication factor will simply change the timescale of growth without altering the structure.

Some representative visualisations are shown in Figure 3. Here we display the fractional neighbour threshold \(\kappa = c/26\), instead of the absolute threshold \(c\), for ease of comprehension. We use default parameters of \(c=20, h=1\) when modelling fitness boost to surface cells; since this leaves a cell with fewer than four neighbours (\(\kappa \leq 0.85\)) with \(\geq 95\%\) of the fitness of an isolated cell; only after that point does fitness begin to decline. 

Varying \(\beta, c, h\) relative to \(\alpha\) (the baseline division rate, kept at \(\alpha = 1\)), and altering the cell-cell adhesion parameter \(\tau\), results in uniformly spherical growth (with the surface rougher or smoother depending on whether \(\tau\) is high or low). The physical reason for this is that if all cells have a non-zero baseline probability of division, then division will occur to some degree or another everywhere in the cluster. Any protrusions that form momentarily at the surface will be evened out by pressure from reproduction beneath the outer layer of cells. Gaps at the surface, corresponding to unoccupied lattice-points, will be filled quickly by newly-produced cells squeezed out from the organoid. This is true even when the division-probability boost is very large (ten times the baseline) and when surface tension is low; in any structure where dividing cells remain in contact with each other, the number of mostly surrounded cells will be much larger than the number of completely isolated cells. Thus most reproduction will happen within the cluster, so instabilities at the surface will not develop and the organoid will remain spherical. In all cases circularity remains above the mutant/wild-type criterion of \(C=0.78\), except for very low \(\tau\) values, where low cell-cell adhesion produces very rough surfaces.

\begin{figure}[H]
 \includegraphics[width=.32\linewidth]{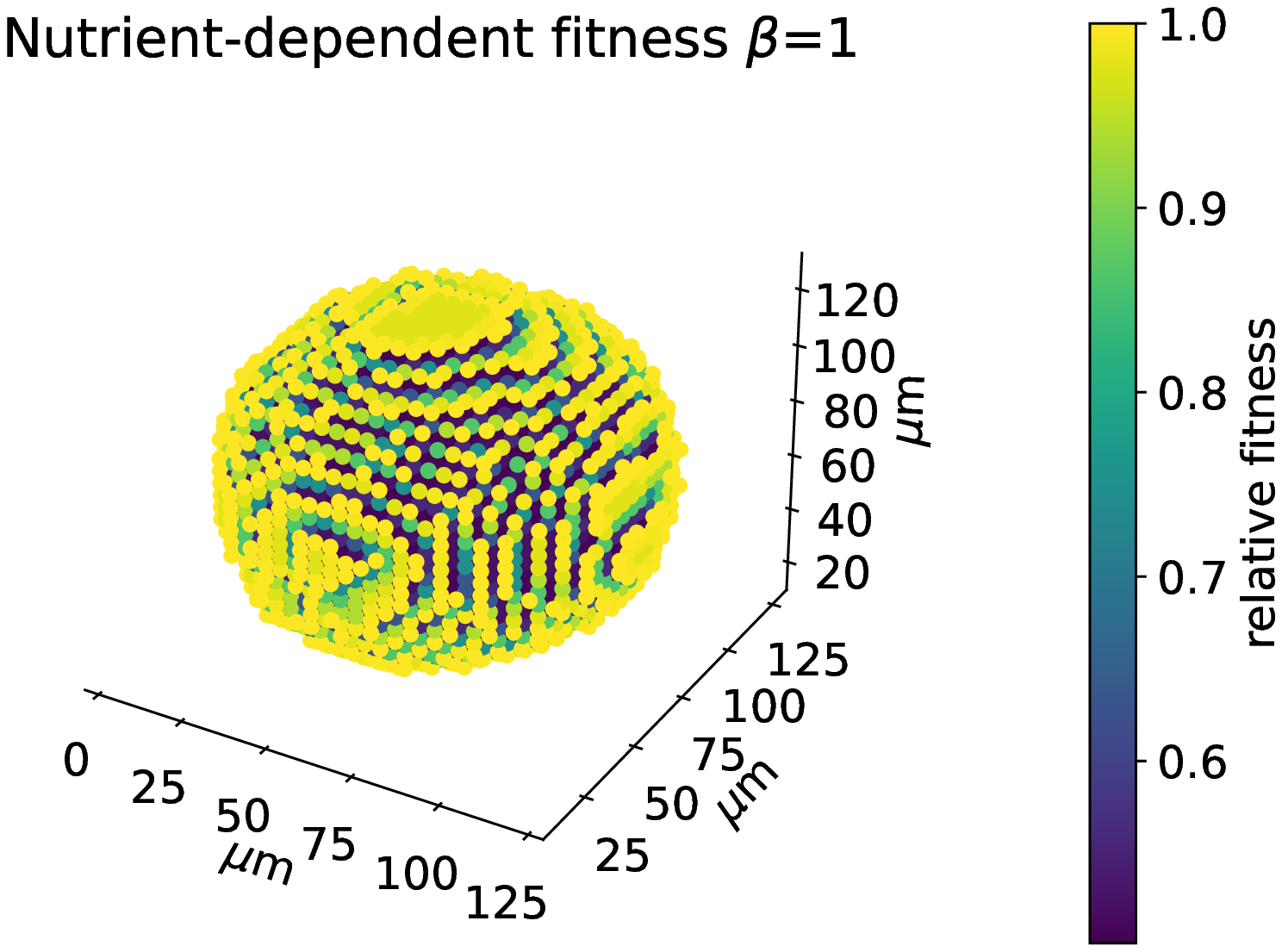}
 \includegraphics[width=.32\linewidth]{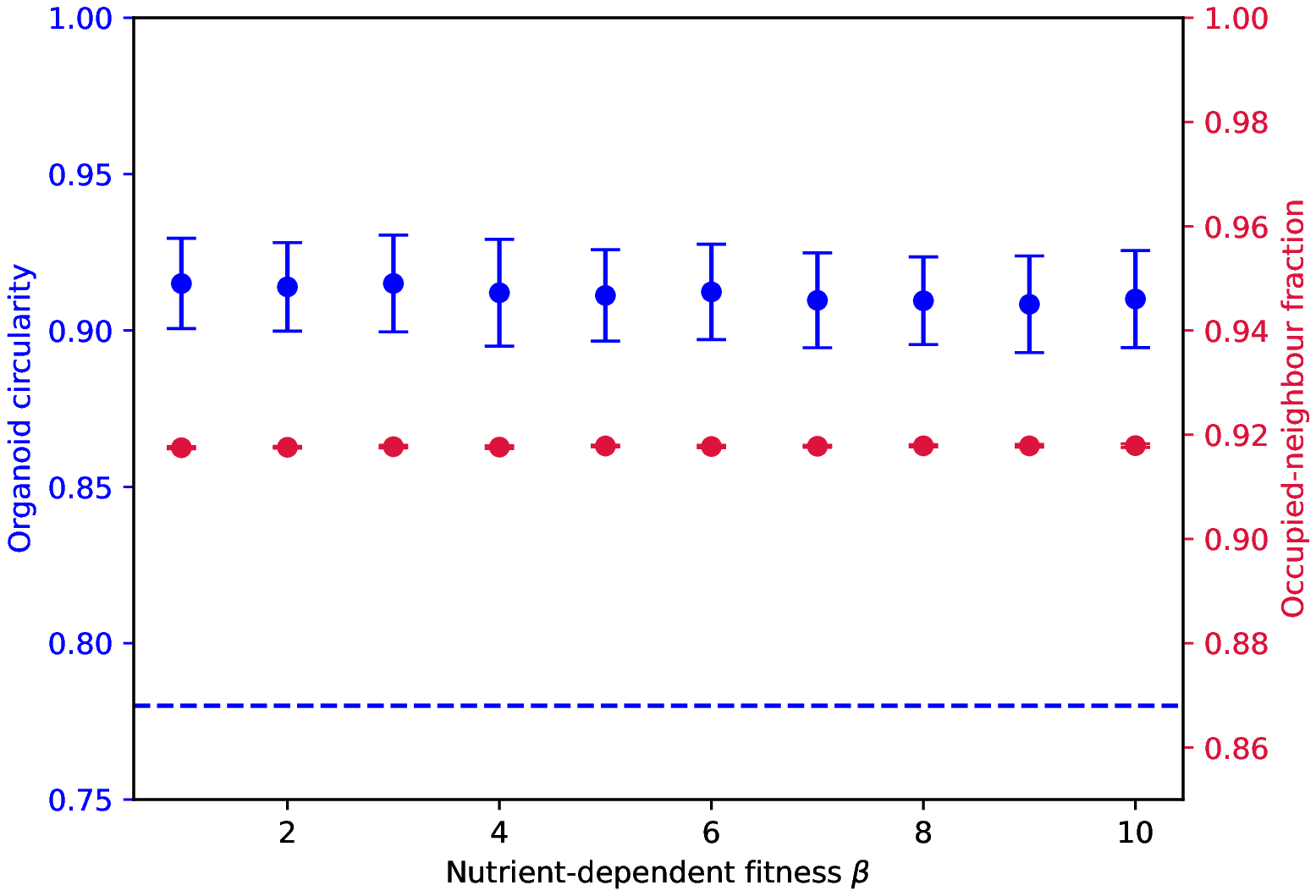}
 \includegraphics[width=.32\linewidth]{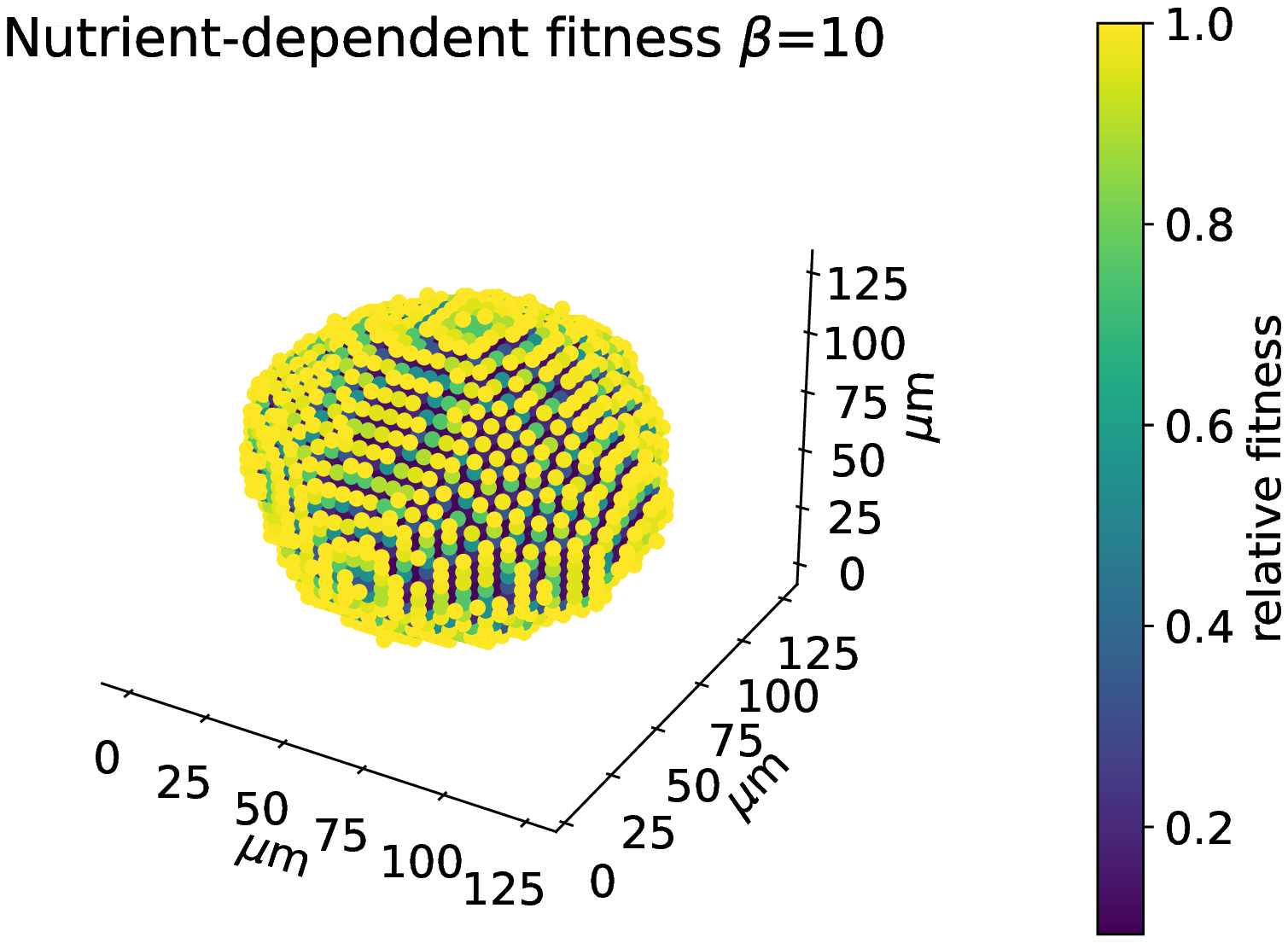}
 \includegraphics[width=.32\linewidth]{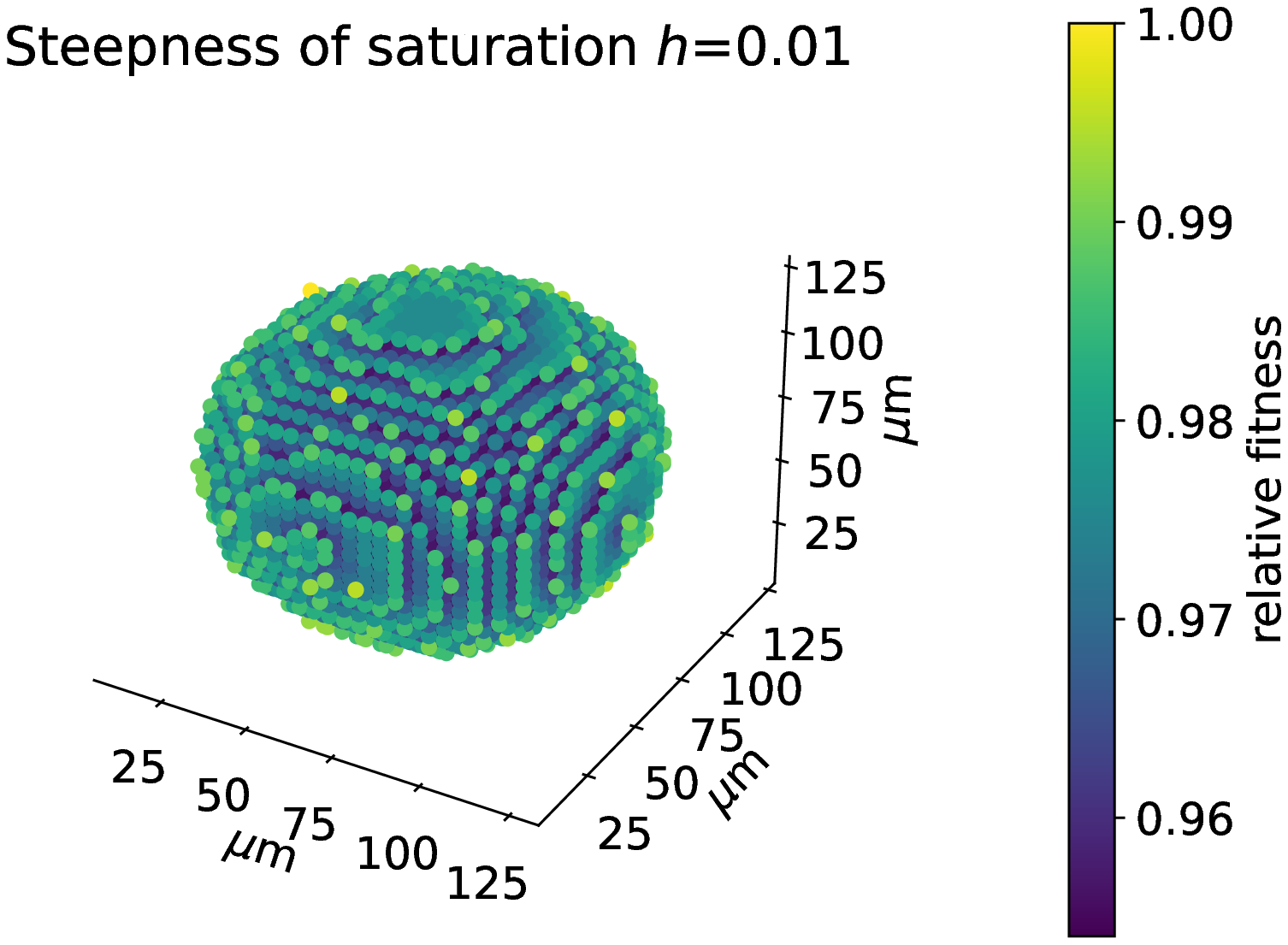}
 \includegraphics[width=.32\linewidth]{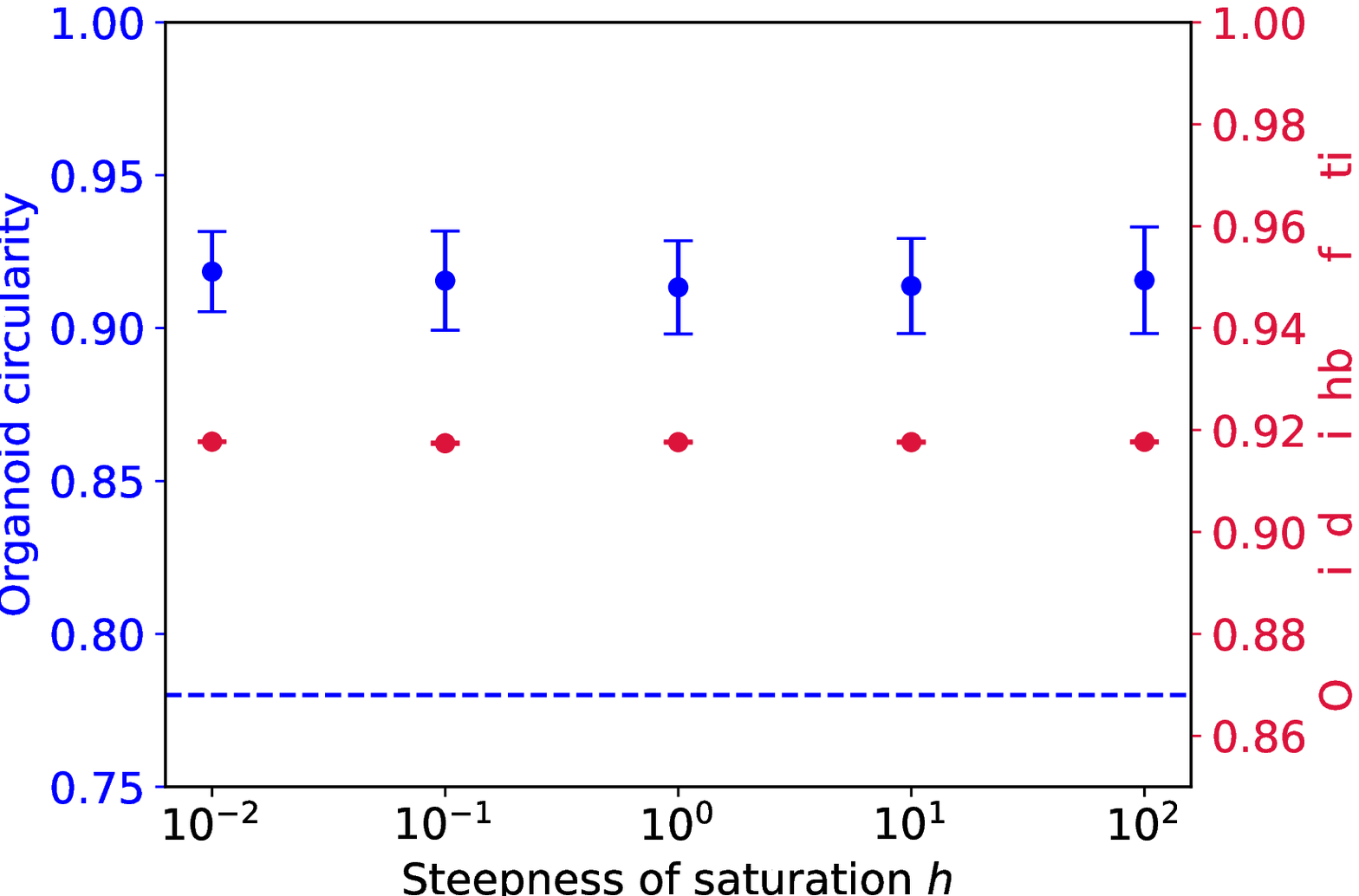}
 \includegraphics[width=.32\linewidth]{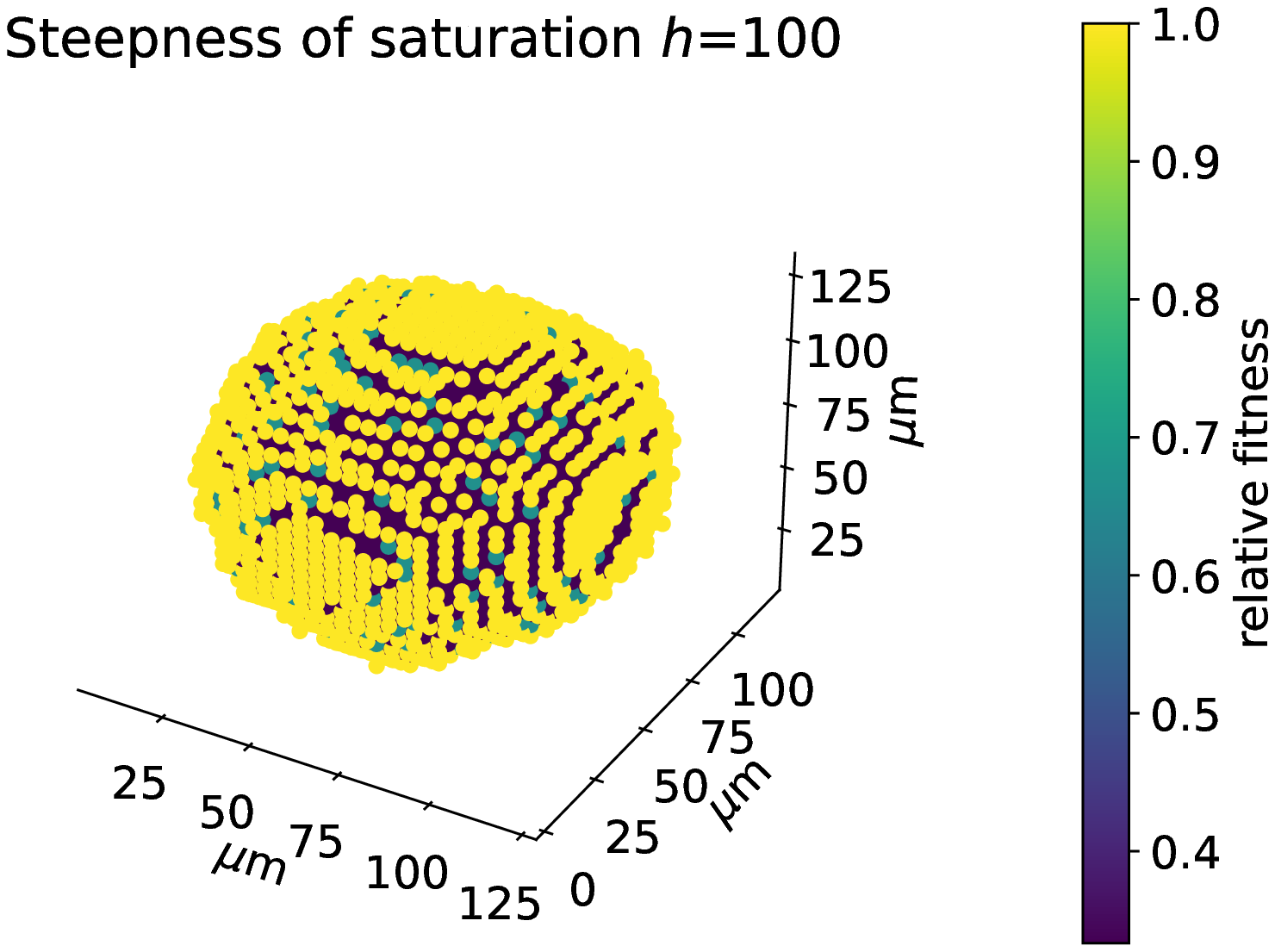}
 \includegraphics[width=.32\linewidth]{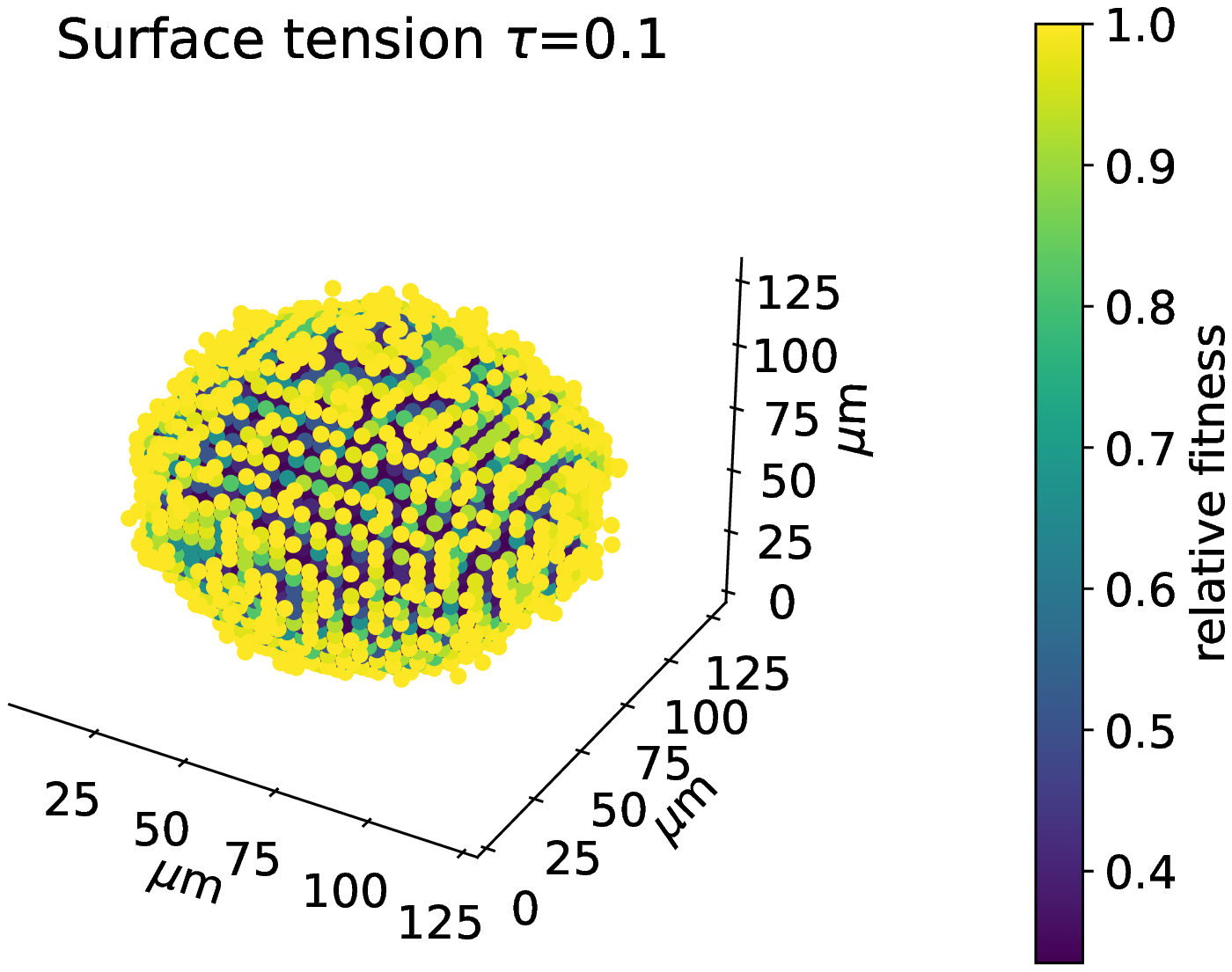}
 \includegraphics[width=.32\linewidth]{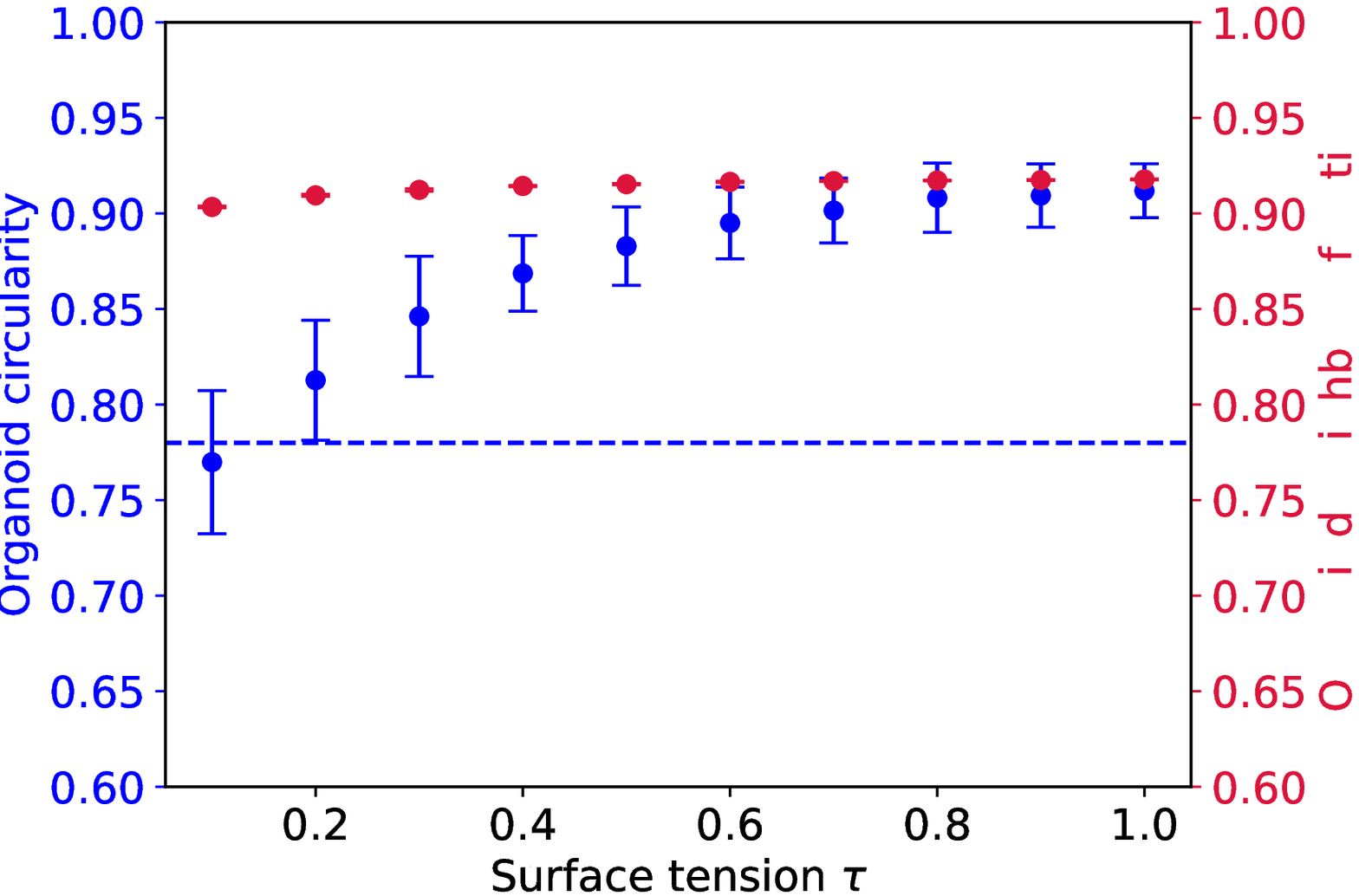}
 \includegraphics[width=.32\linewidth]{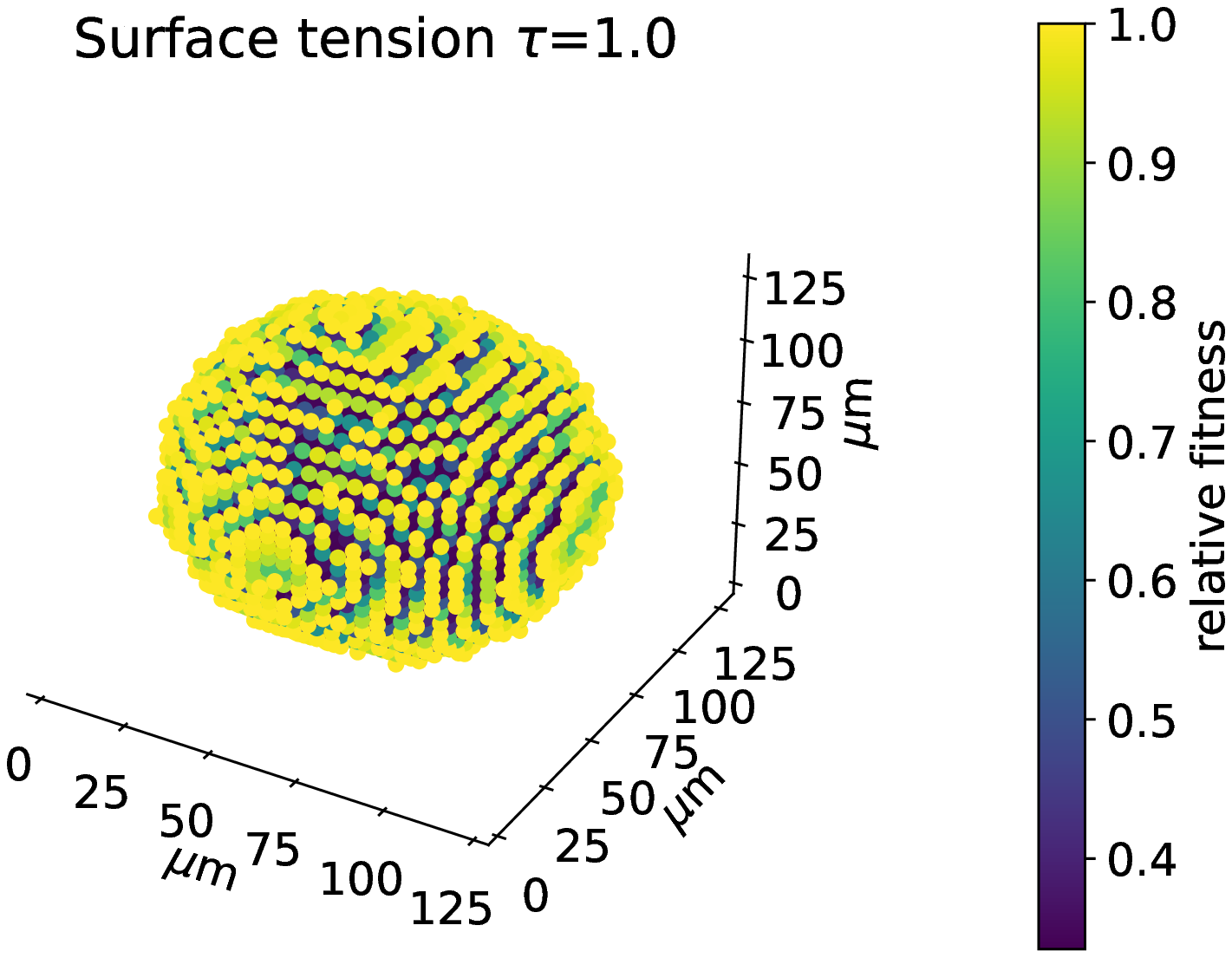}
 \includegraphics[width=.32\linewidth]{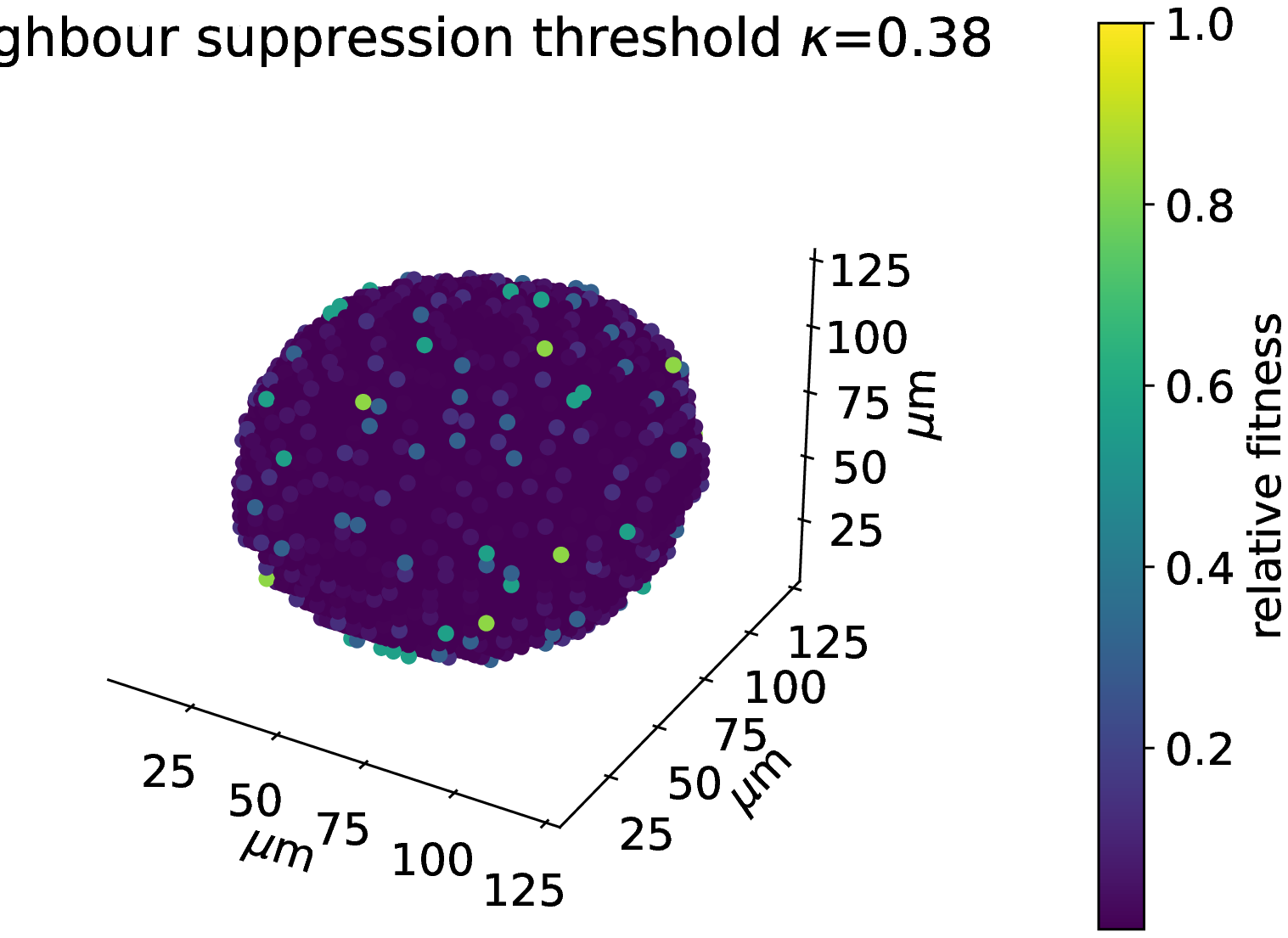}
 \includegraphics[width=.32\linewidth]{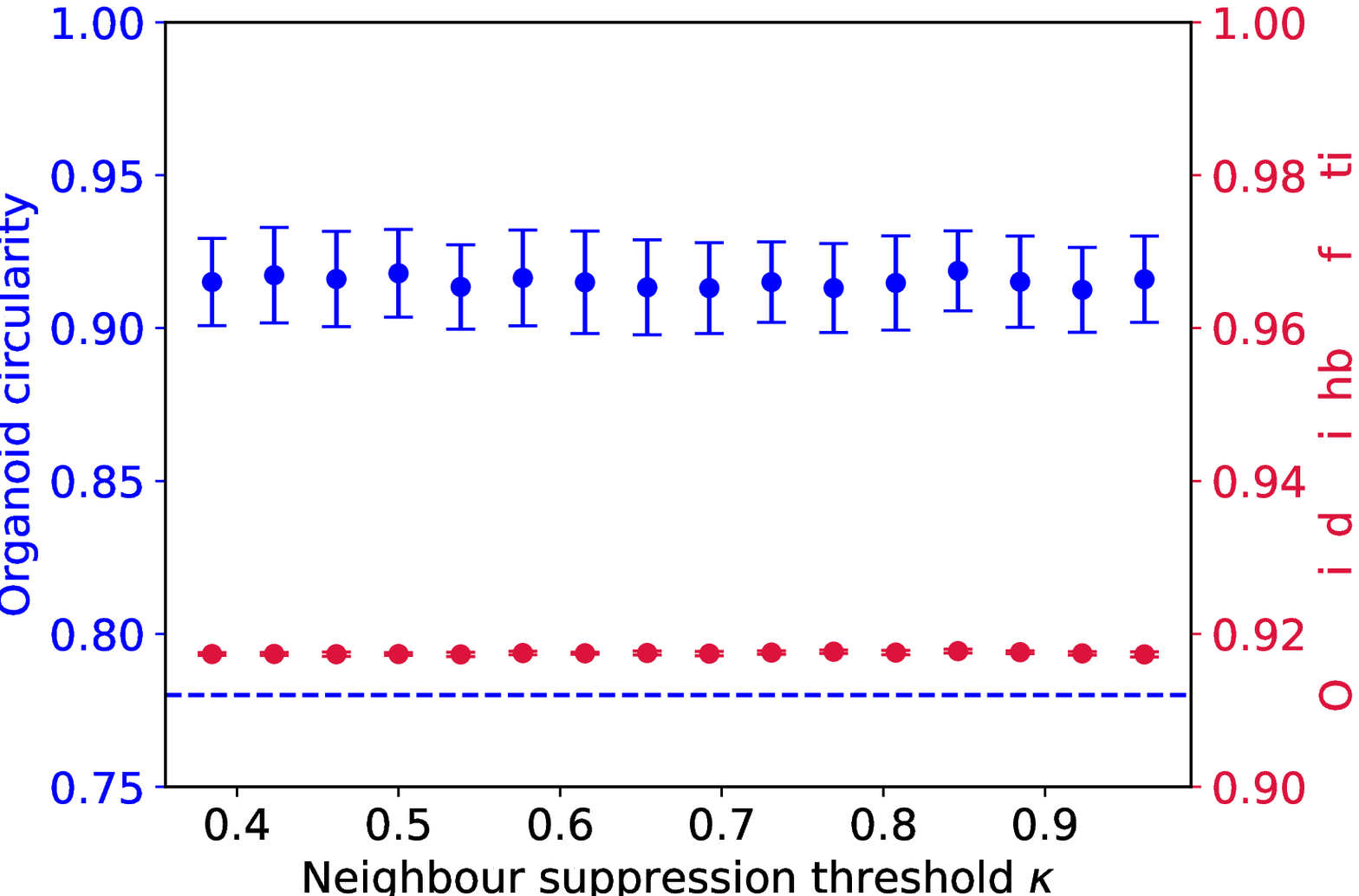}
 \includegraphics[width=.32\linewidth]{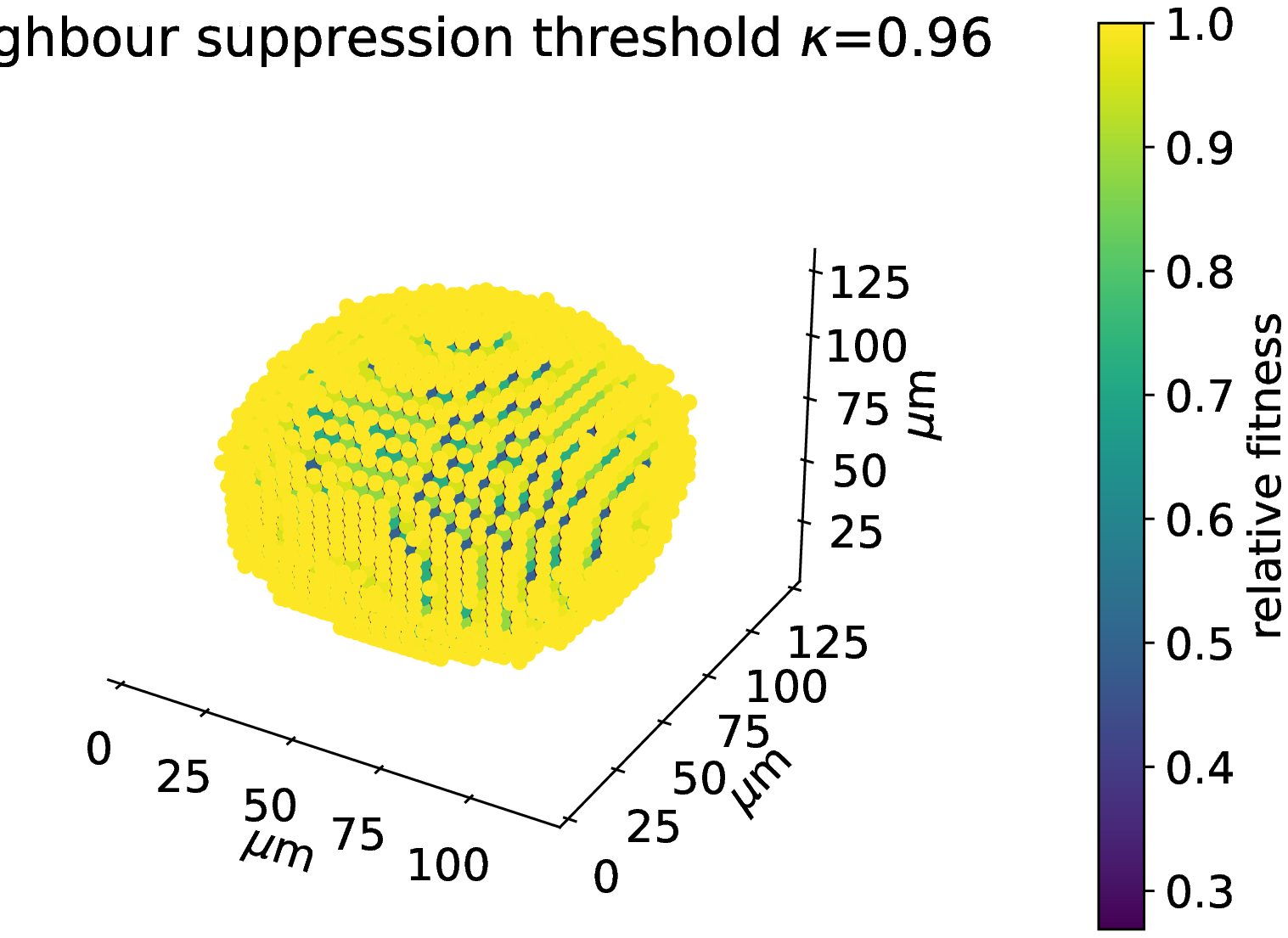}
  \caption{Illustrations of the effect of varying (top row) `boost' \(\beta\), (second row) steepness \(h\), (third row) cell-cell adhesion \(\tau\), and (fourth row) fractional threshold \(\kappa\). Default parameters are \(\beta=2, c=20 
 (\kappa = 0.77), h=1, \tau=1\). Each graph shows organoid circularity (blue) and the average fraction of an organoid-cell's neighbourhood lattice-points which are occupied (red); a completely surrounded cell has occupied-neighbour fraction 1, and an isolated one has occupied-neighbour fraction 0. These measurements are averaged over 10 simulated organoids and (in the case of circularity) 9 different viewpoint angles of each organoid. Error bars show standard deviations. The circularity threshold below which an organoid is `mutant-like', \(C=0.78\), is indicated with the dotted blue line. Beside each graph, illustrations of the effects of extreme values on organoid structure are included. Here the colour of each cell indicates its relative reproductive fitness (RRF), scaled so the fittest cell has an RRF of 1. Each simulation is run until the first timestep in which the cluster contains more than 10,000 cells. All axes are in \(\mu m\); here 10,000 cells gives an experimentally realistic diameter of \(100 \mu m\).}
\end{figure}

\subsubsection{Anchorage-dependent growth}

Another plausible hypothesis is `anchorage-dependent growth' \cite{Ruoslahti1994}, whereby cells require adhesion to extra-cellular matrix (ECM), and thus contact with other cells, in order to progress through the cell cycle. This suggests that division probability increases monotonically with the number of neighbouring cells. This motivates a division probability function of the form

\[p_{anc}(\alpha, \beta, h, c, n, dT) = p_{abs}(\alpha, \beta, -h, c, n, dT) =  \left(\alpha + \frac{\beta}{1+e^{h(c-(26-n))}}\right)dT \]where all parameters have the same meaning as before. Now \(c\) is the threshold number of occupied neighbours at which a cell's proliferative capacity \textit{increases}, and this capacity \textit{declines} with \(n\), a cell's number of empty neighbouring lattice-points. We find that this leads to spherical growth since cells benefit from being as close to one another as possible, with protruding cells at a disadvantage (see Figure 4).

\begin{figure}[H]
    \includegraphics[width=.49\linewidth]{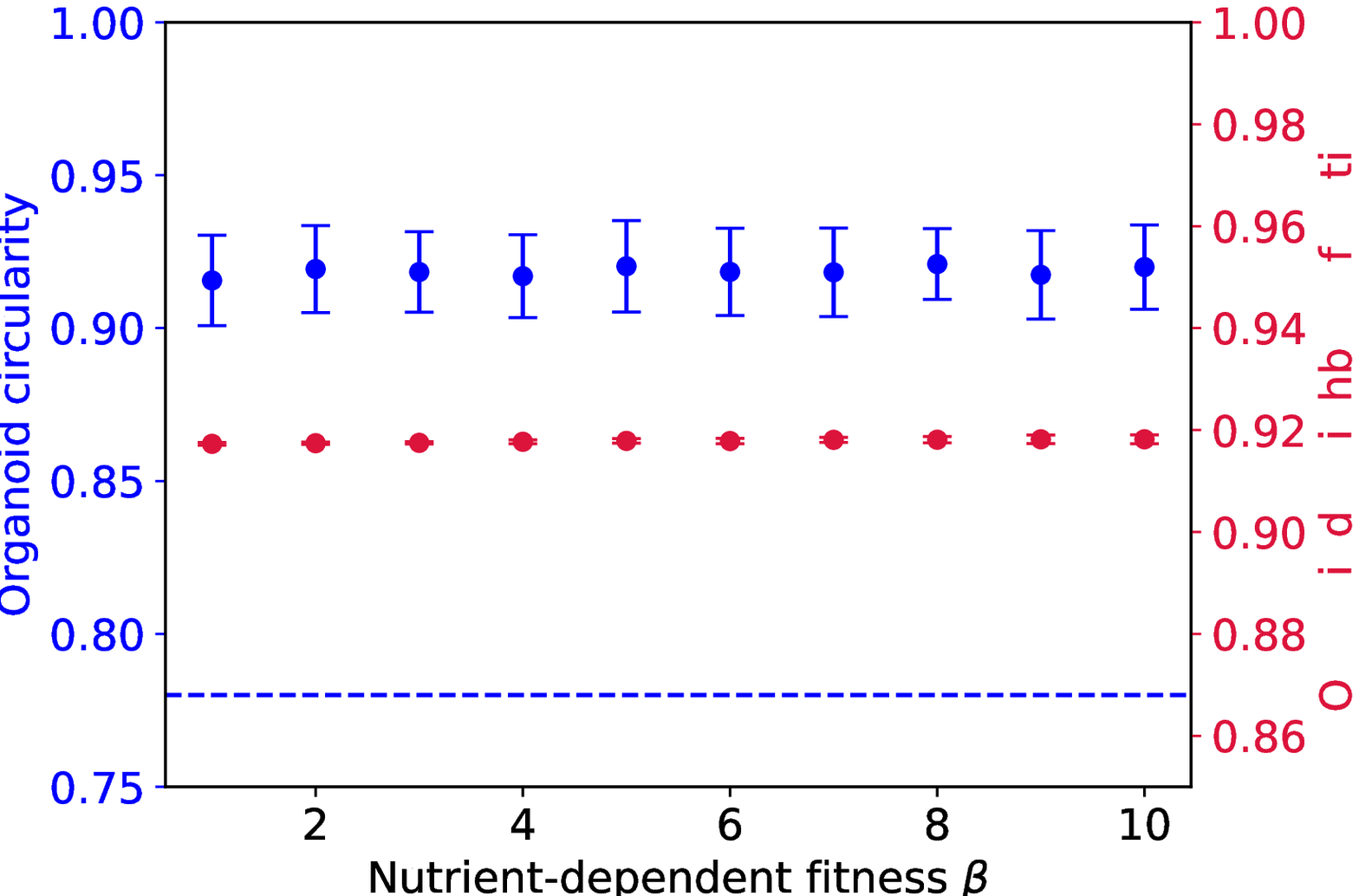}
    \includegraphics[width=.49\linewidth]{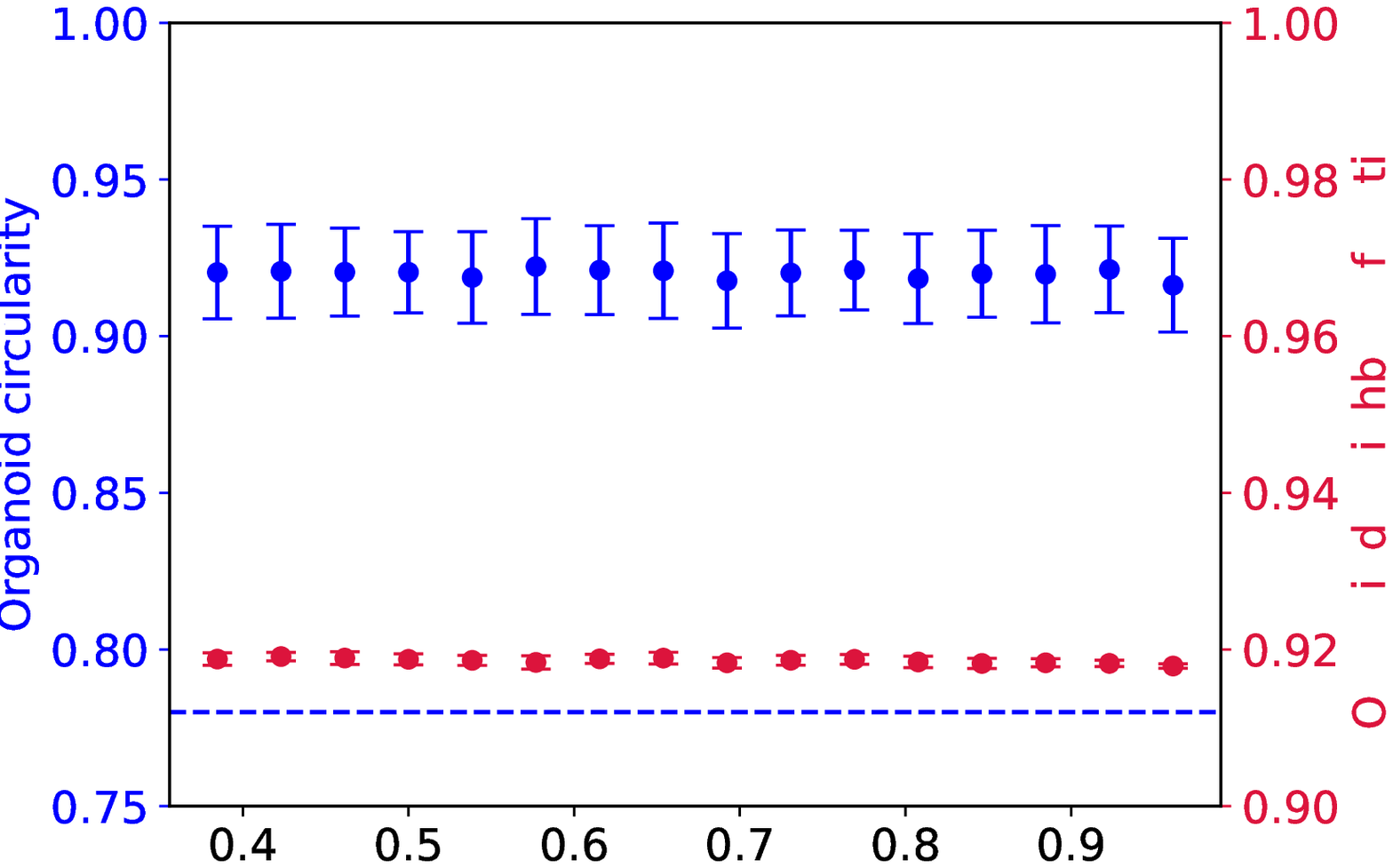}
  \caption{Illustrations of the effect of varying (left) `boost' \(\beta\) and (right) fractional threshold \(\kappa\) on organoid structure under a hypothesis of anchorage-dependent growth. Default parameters are \(\beta=10, c=20 (\kappa=0.77), h=1, \tau=1\). Secondary spheroids are never produced. All annotations are as Figure 3.}
\end{figure}

\subsection{High levels of differentiation combined with efficient use of nutrients by mutant cells can result in `budding structures'}
A further testable hypothesis is differentiation-based growth. A dividing progenitor cell might produce, for example, a progenitor cell (which can reproduce in turn) or a differentiated cell (which cannot). Could differences in the division patterns of progenitor cells create budding structures?

To test this, we make a slight alteration to our framework. We use a division function of the form \( p_{abs}\), assuming some fitness boost to surface cells, but further divide cells into two types: `active' and `inactive'. Both cells can be pushed aside when other cells reproduce, but only active cells can divide. When an active cell divides, the new cell is active with probability \(1-q\), and inactive with probability \(q\). We also implement a change during the adjustment step, to prevent newly-produced cells from ending up very far from their mother cells as soon as they are produced. We keep track of the type of cell last added to any lattice point. When a cell is moved from that point, we make sure to keep at least one cell of that type there after the movement step. The moved cell is chosen at random from those remaining. For example, if a point initially contains one active cell, and has an inactive cell moved to it during the adjustment step (so that it is double-occupied), then in the next step, the active cell will be moved. This ensures that cells are not propagated too far through the cluster during a single adjustment step.

We see that high levels of differentiation enhances the structural effects of absorption-dependent fitness to produce `bubbling growth'. The combination of these two mechanisms is sufficient to limit division to a small number of cells almost entirely located on the surface. 

In the section below we consider other model systems which can produce organoids below the `mutant-like' circularity threshold. Whilst they differ in their physical motivation, all produce this same surface-localisation effect.

\begin{figure}[H]
    \includegraphics[width=0.32\linewidth]{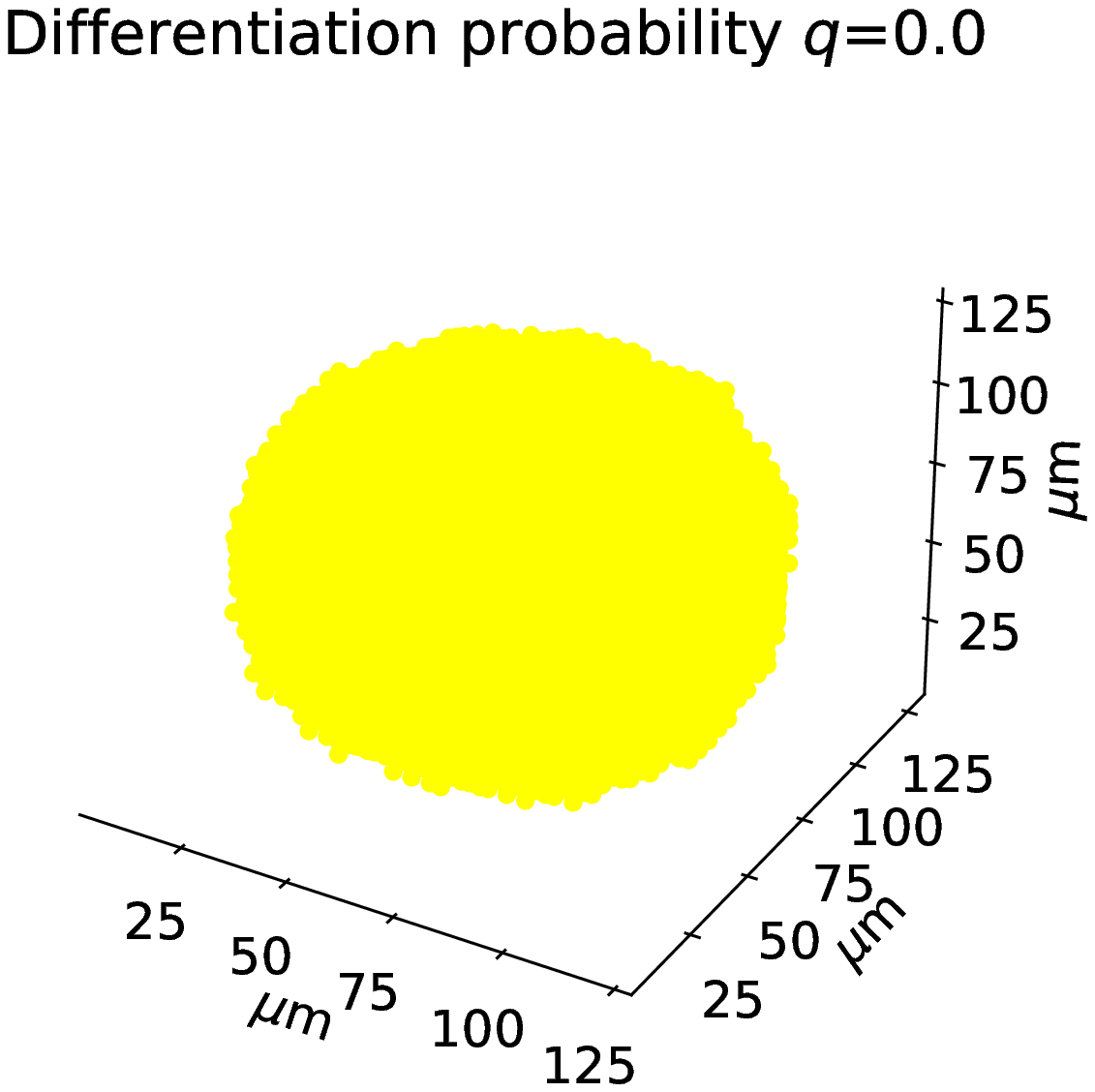}
   \includegraphics[width=0.32\linewidth]{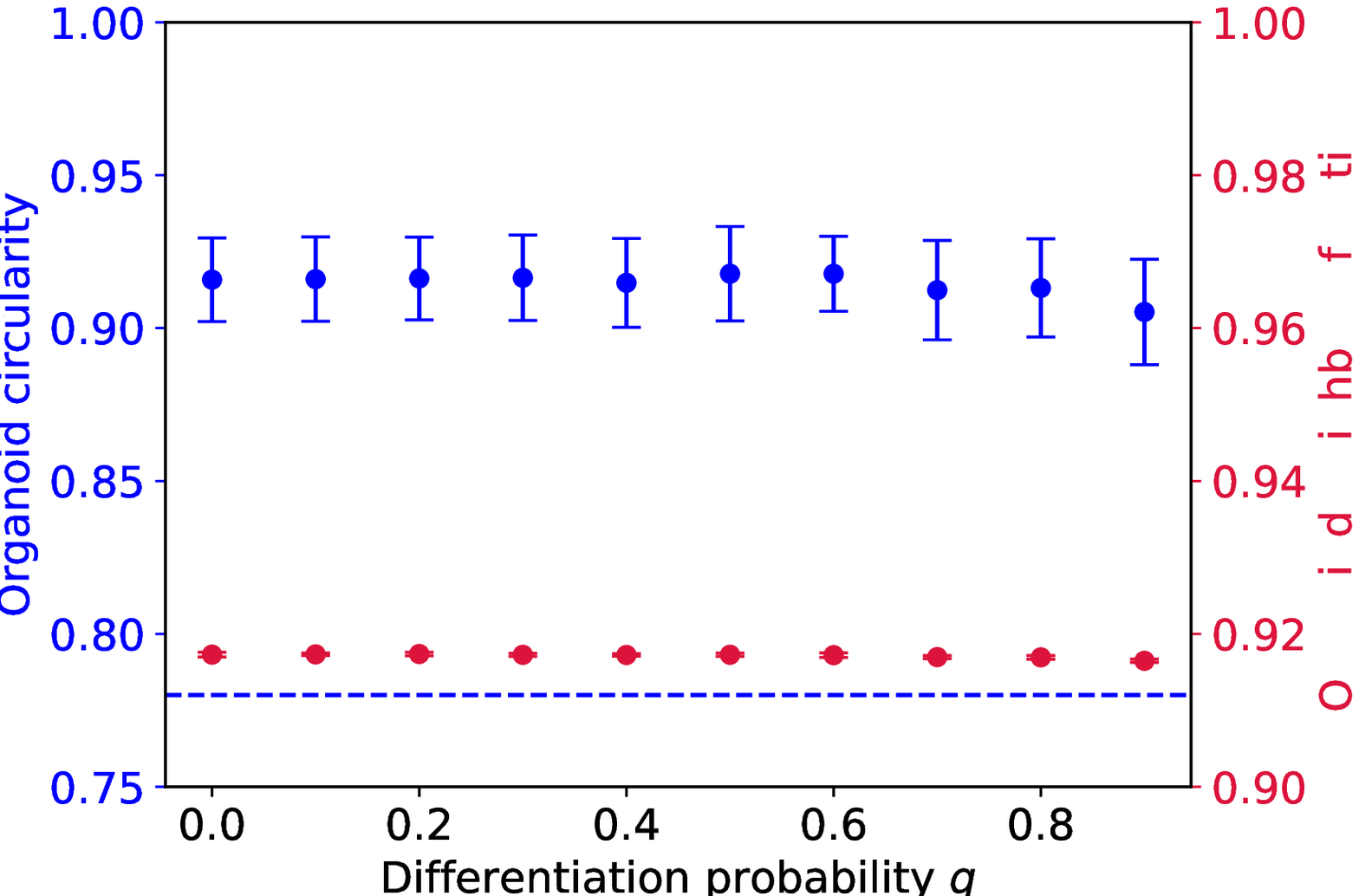}
   \includegraphics[width=0.32\linewidth]{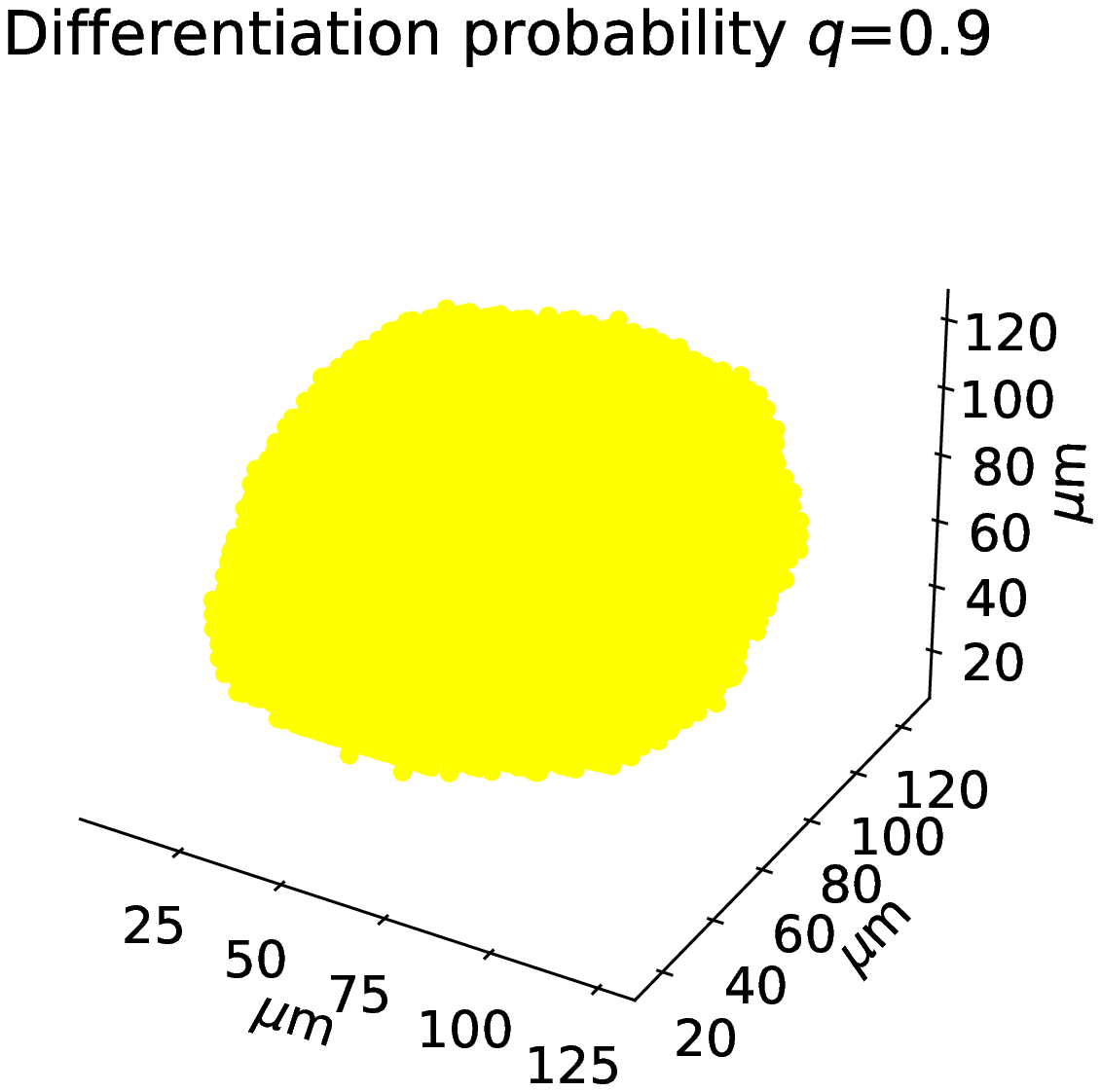}
   \includegraphics[width=0.32\linewidth]{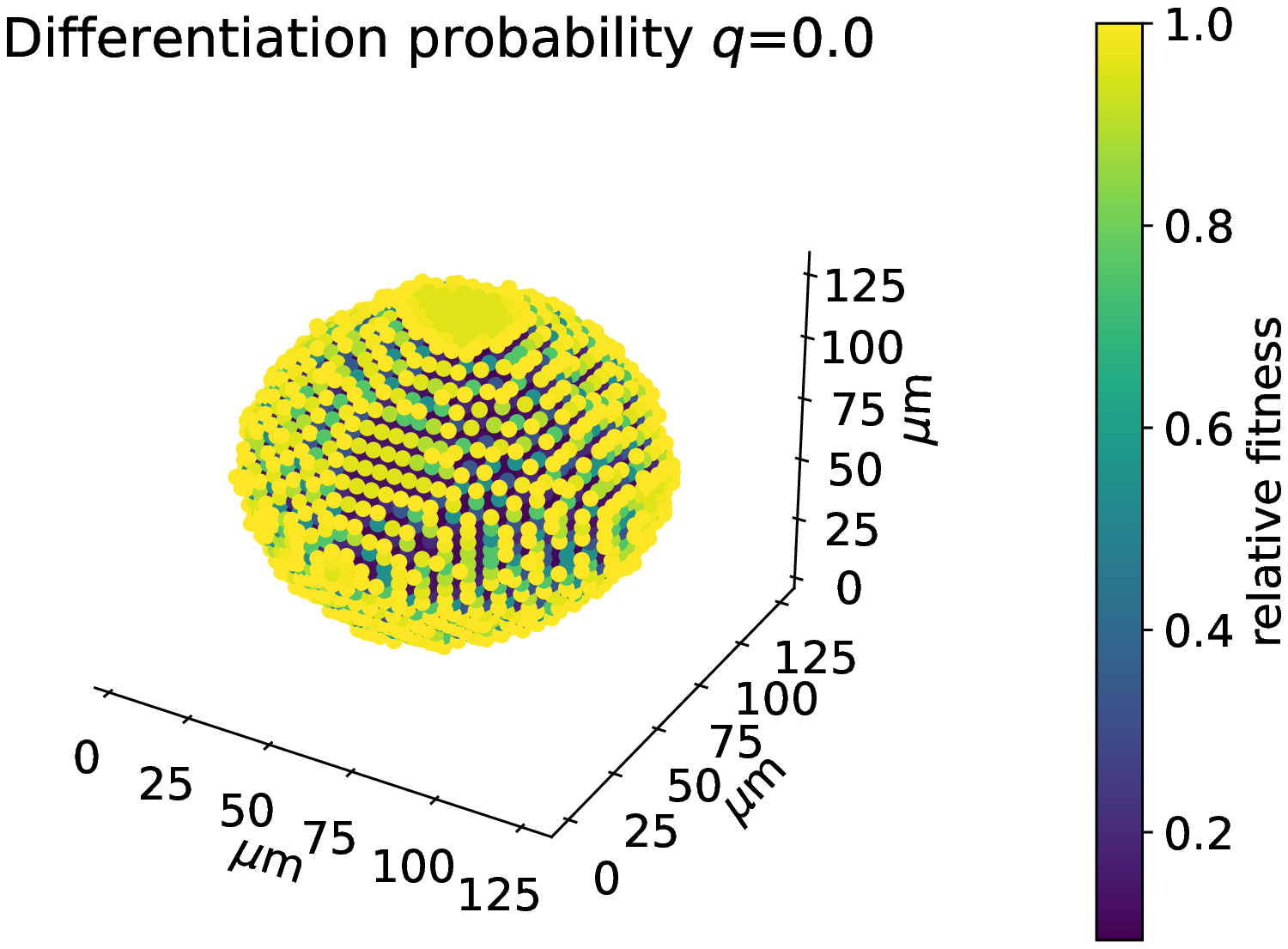}
   \includegraphics[width=0.32\linewidth]{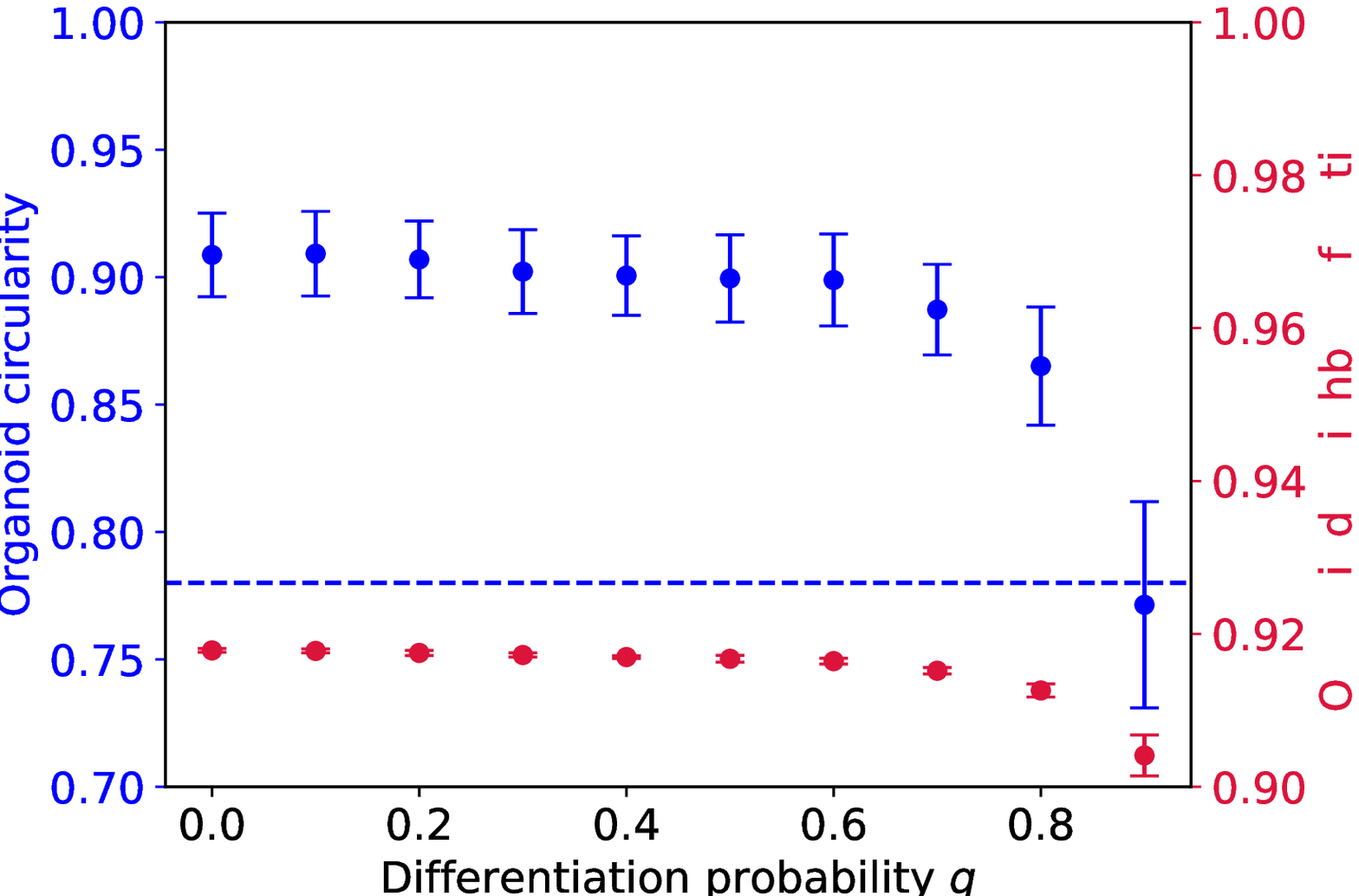}
   \includegraphics[width=0.32\linewidth]{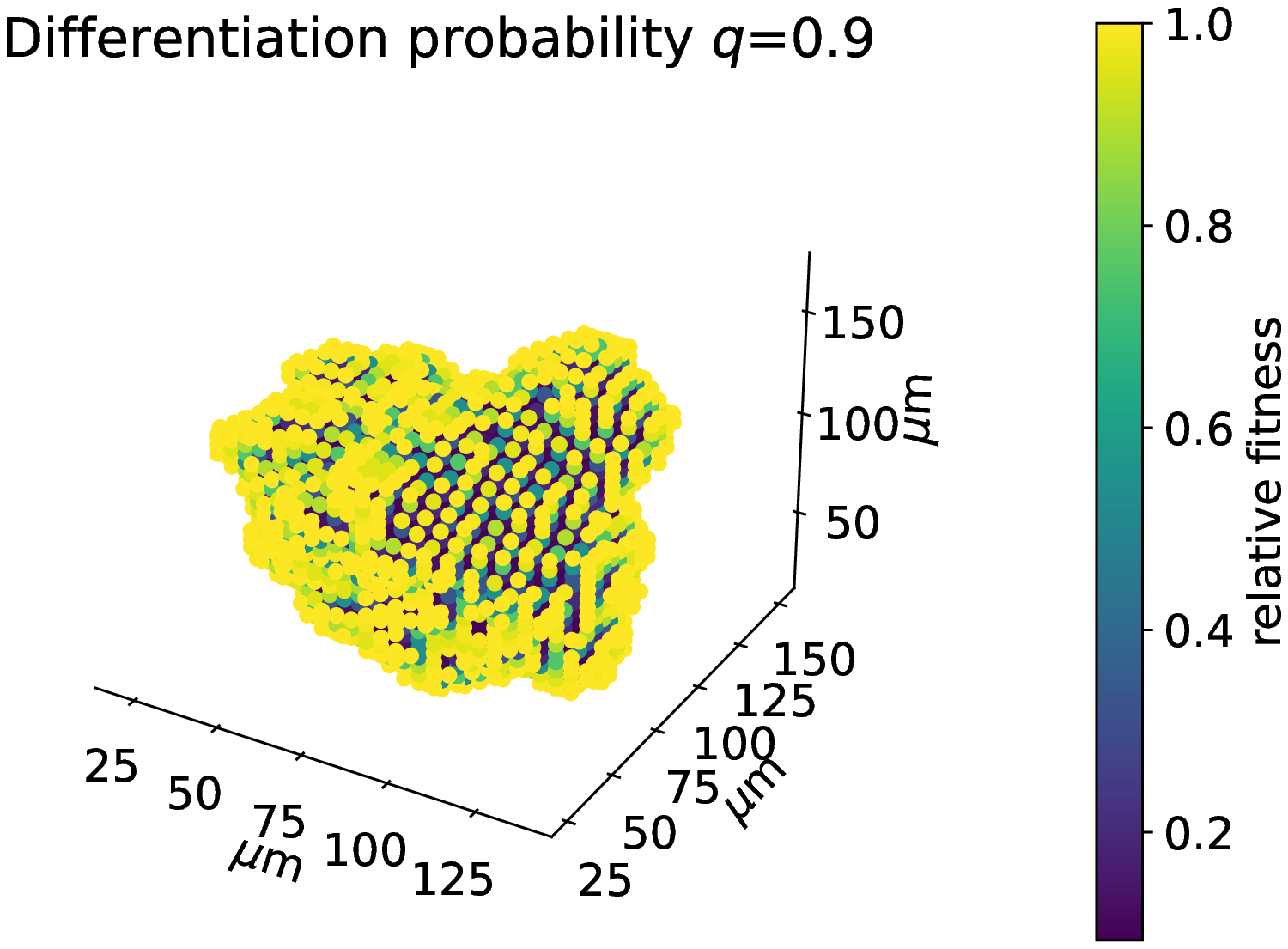}
  \caption{The effect of varying differentiation probability on organoid structure: with (left) \(\beta=0\), and (right) with \(\beta=10, c=20 (\kappa=0.77), h=1, \tau=1\). In all experiments \(\tau=1.0, \alpha=1\). All experiments are run to 10,000 cells. Here colour indicates the fitness a cell \textit{would} have were it active; the positions of active and inactive cells are not shown.}
\end{figure}

\subsection{Strong neighbour suppression is sufficient to produce mutant organoid structures}

The simulations above establish that so long as surrounded cells are able to reproduce, clusters will remain largely spherical. What happens if surrounded cells are completely prevented from dividing through means other than differentiation? To test this hypothesis, we use the division probability function 

\[p_{\text{NS}}(n, dT) = p_{abs}(0, \beta, h, c, n, dT) = \frac{\beta dT}{1+e^{-h(c-(26-n))}} \]
which is simply \(p_{abs}(n, dT)\) with \(\alpha = 0\).  Here cells must have empty neighbour-points in order to reproduce; the baseline division probability of surrounded cells is almost zero. This may happen for a couple of mechanistic reasons. If cells are sufficiently densely packed that nutrients cannot reach cells at the centre of the cluster, we might expect them to stop dividing. This would only happen when cells are completely surrounded, however, and so would correspond to a high threshold \(c\). If cells were stopped from dividing when not completely surrounded, however, we might attribute this to active inhibition by neighbouring cells.

Using this growth mechanism, sustained protrusions can develop, and we observe structures which look remarkably similar to those seen in experiments (Figure 6). Mutant-like circularity is established at a threshold of \(c=13\) or \(\kappa = 0.5\) and a saturation steepness of \(h=1\), regardless of the level of cell-cell adhesion. This corresponds to a division probability which is saturated for a cell with fewer than 11 neighbours (where 42\% of its surface is covered) and close to zero for a cell which has more than 15 (where 58\% of its surface is covered). This low threshold is worth noting: cells that are still very much in contact with the nutrient-rich Matrigel cannot divide. These extremely high levels of neighbour inhibition suggest either that local competition for nutrients is so strong that even surface cells can be deprived of resources by their neighbours, or else that mutant cells actively inhibit each other's growth through mechanotransduction. Both explanations take the same mathematical form and so are permitted by this model.

 Decreasing the steepness \(h\) means that the benefit of neighbouring empty lattice points (i.e. the inhibitory effect of neighbour-neighbour signalling) becomes approximately linear, which allows surrounded cells to divide, leading to spherical growth. In order to generate secondary spheroids, the benefit to surface cells must be steeply nonlinear, so the behaviour of these cells is a `switchlike' function of their circumstances. Surface-dependent growth requires the `surface' to be distinctly and sharply advantaged, i.e. for the proliferative capacity of a cell to increase significantly when it has below a certain number of neighbours. 

Cross-sectional analysis of the simulation shows that these organoids grow as solid structures, without `air pockets', which also matches experimental observation (when grown organoids are removed from the Matrigel and analysed at the end of the experiment, they are found to be solid).

\begin{figure}[H]
    \includegraphics[width=.32\linewidth]{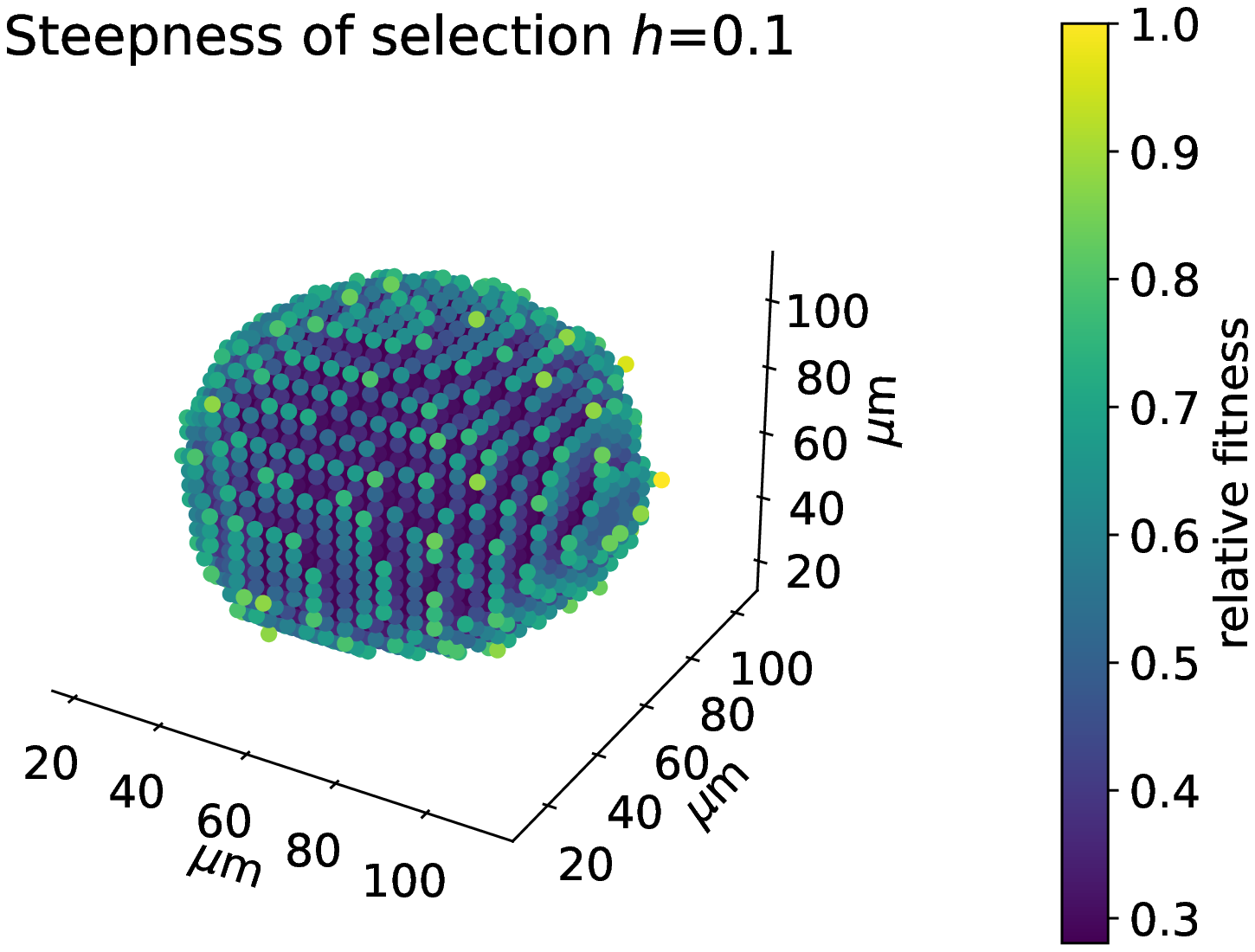}
    \includegraphics[width=.32\linewidth]{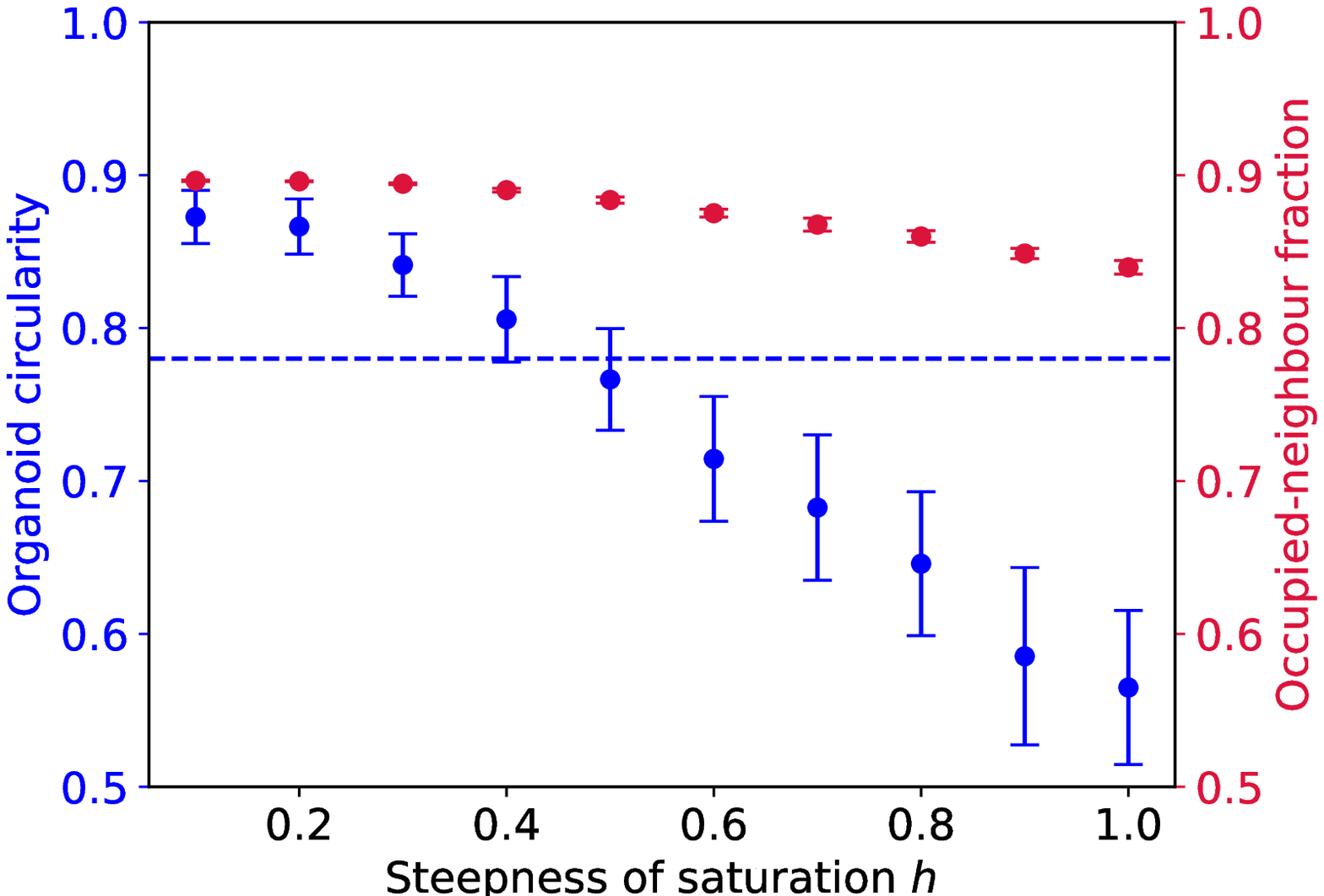}
    \includegraphics[width=.32\linewidth]{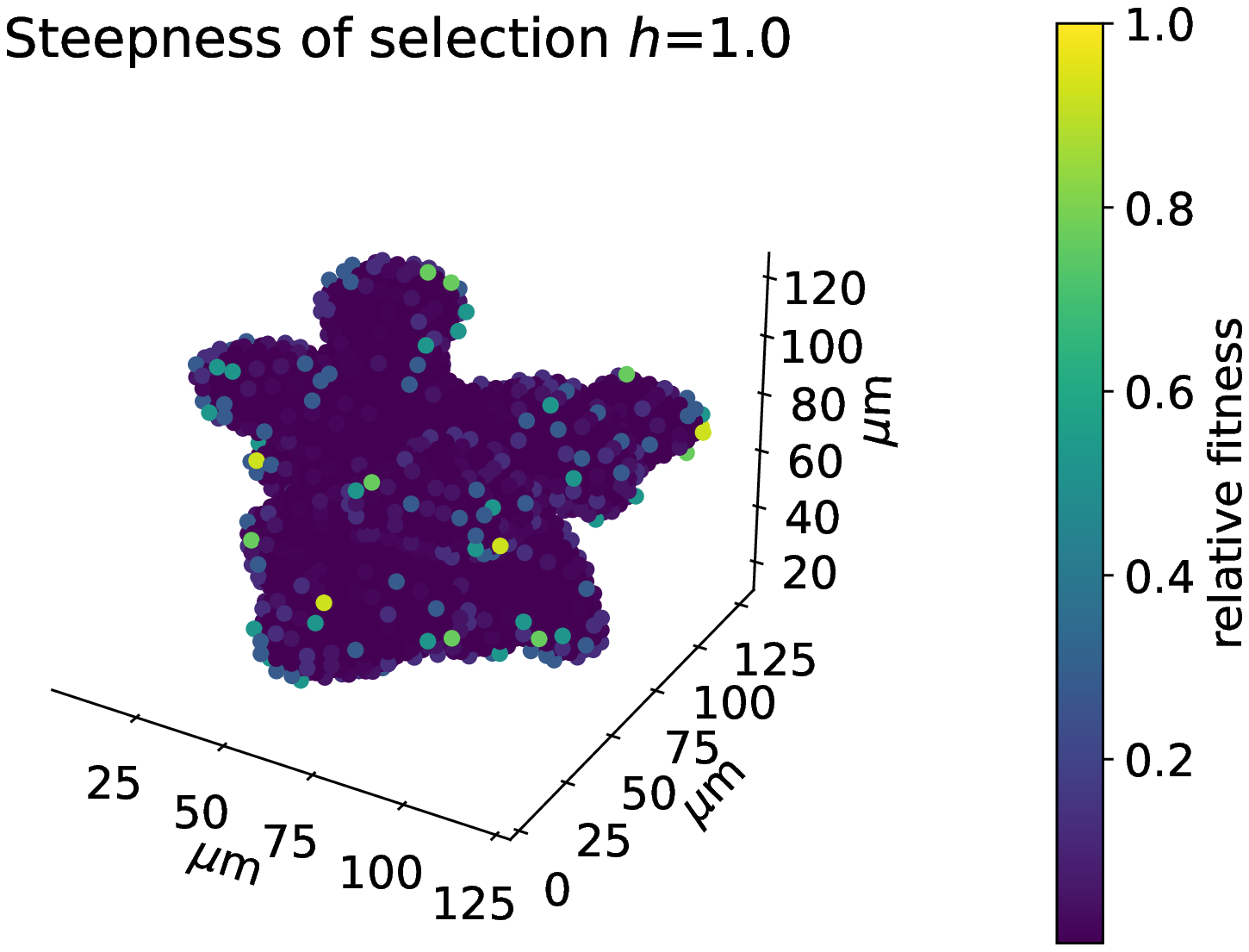}
    \includegraphics[width=.32\linewidth]{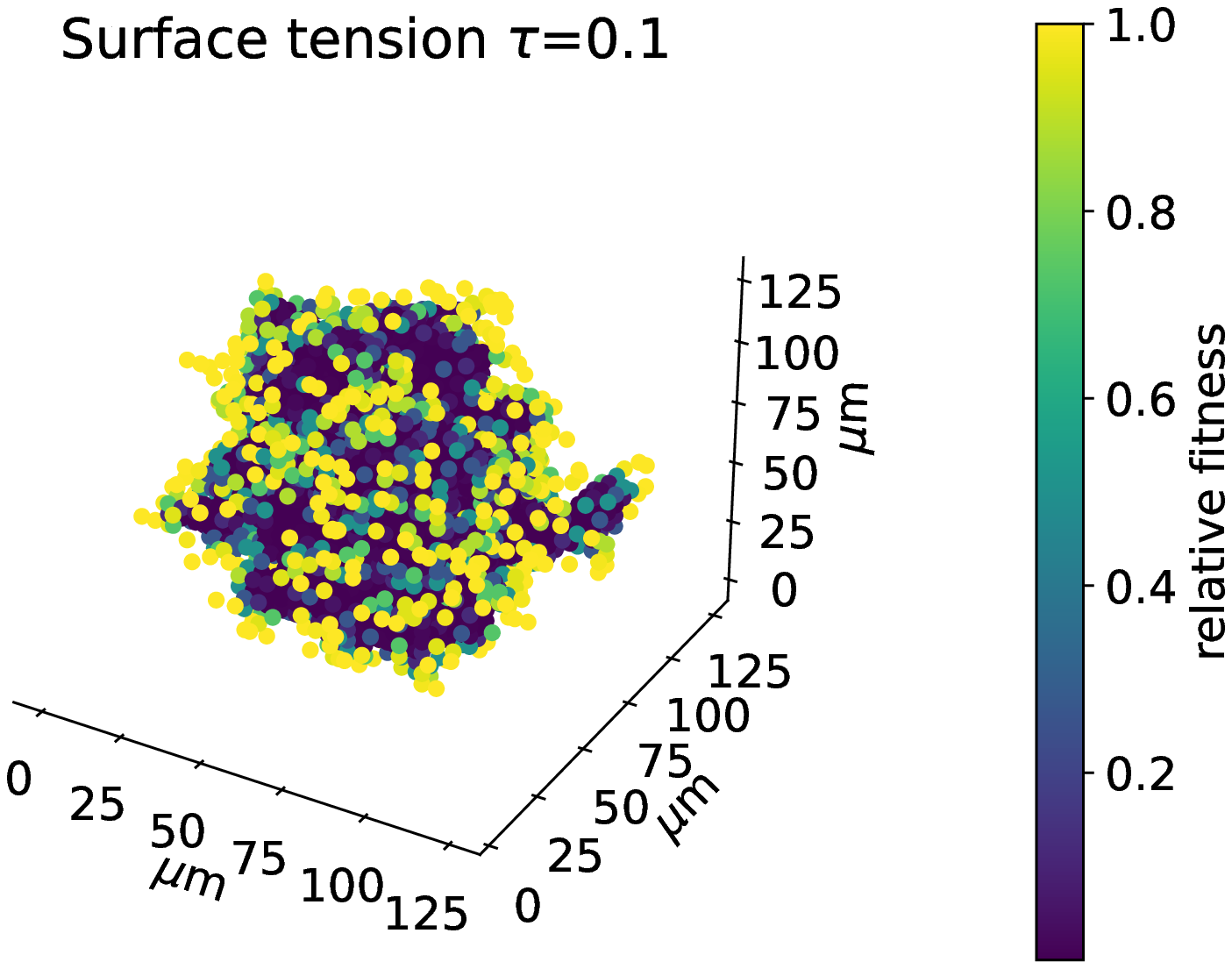}
    \includegraphics[width=.32\linewidth]{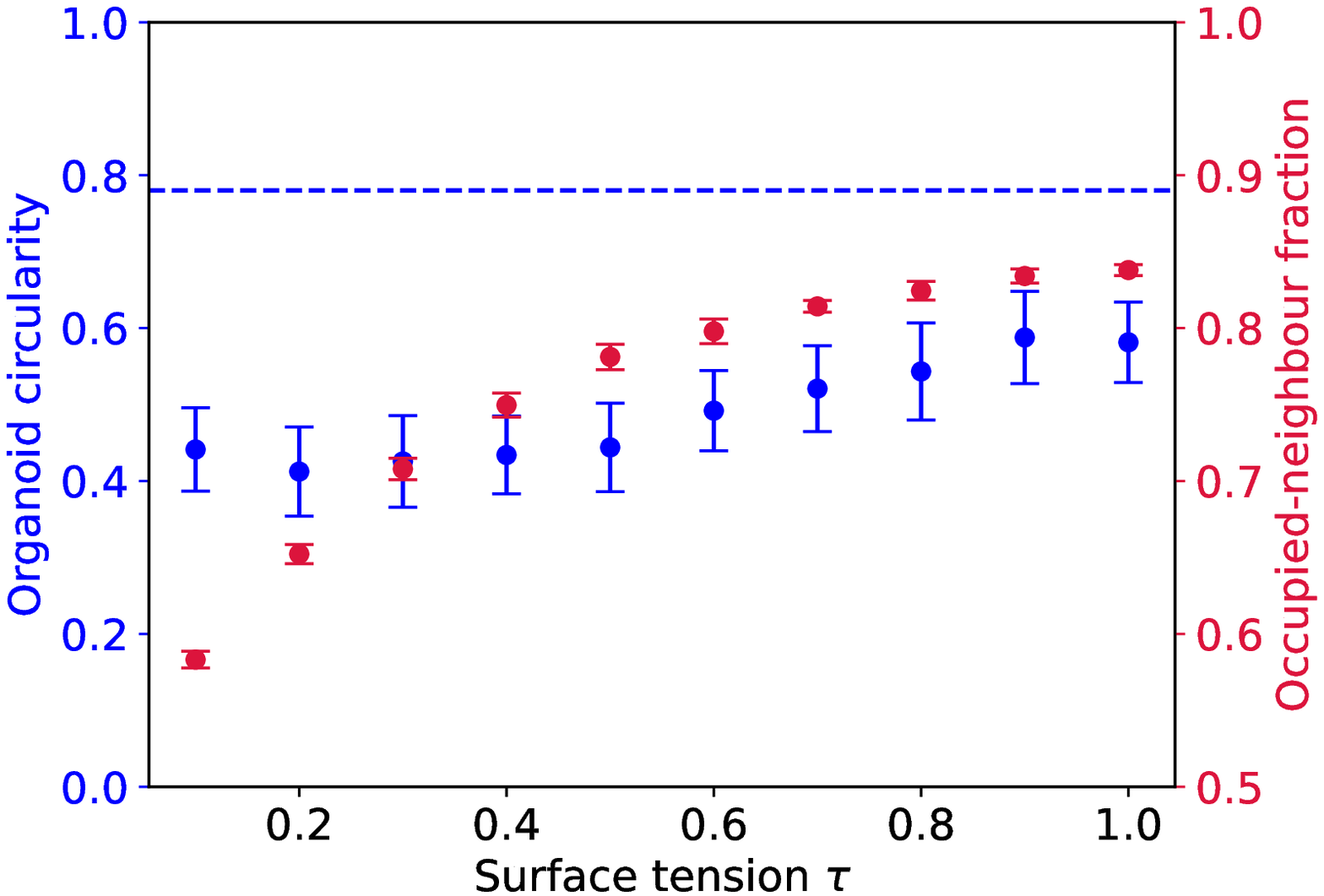}
    \includegraphics[width=.32\linewidth]{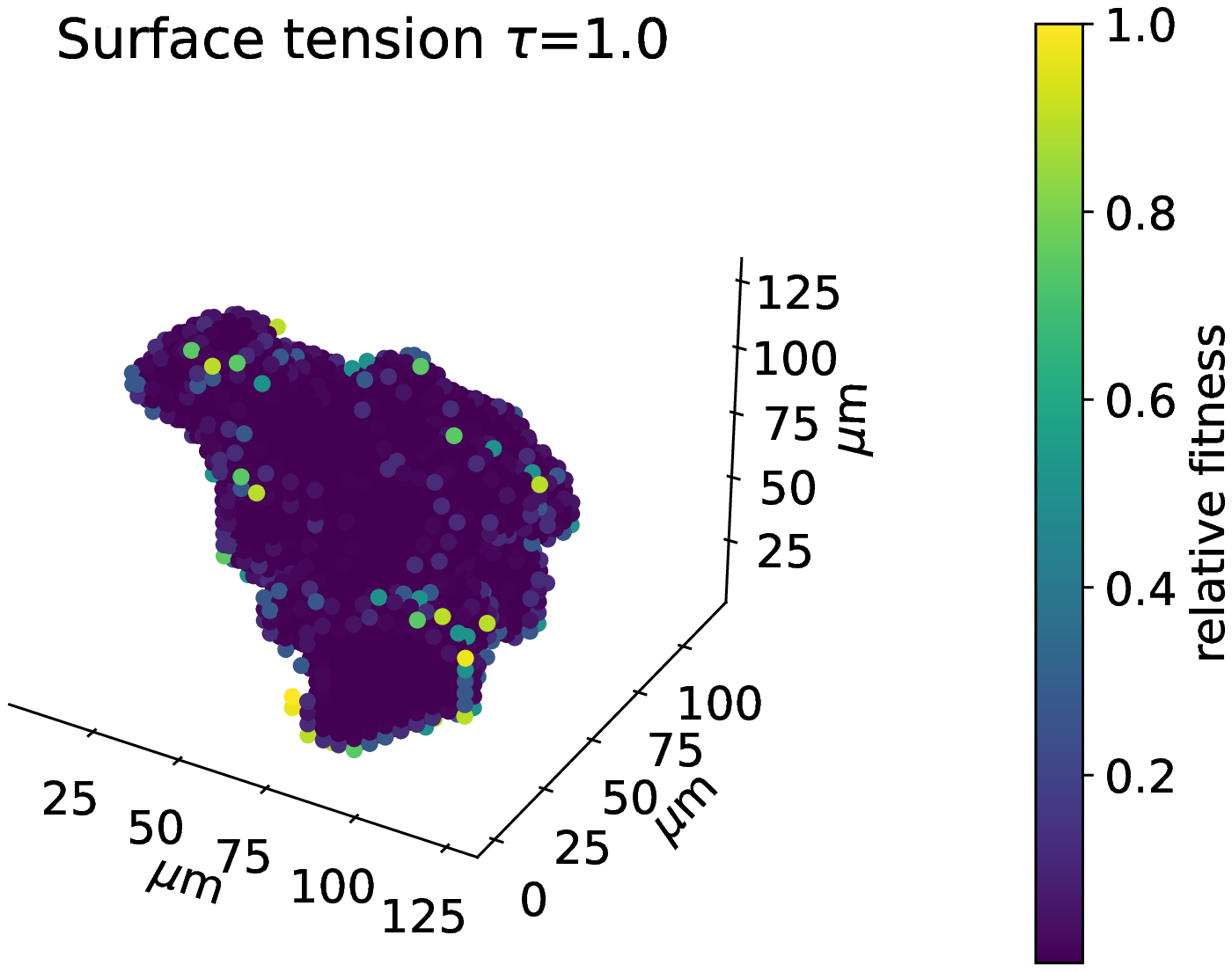}
    \includegraphics[width=.32\linewidth]{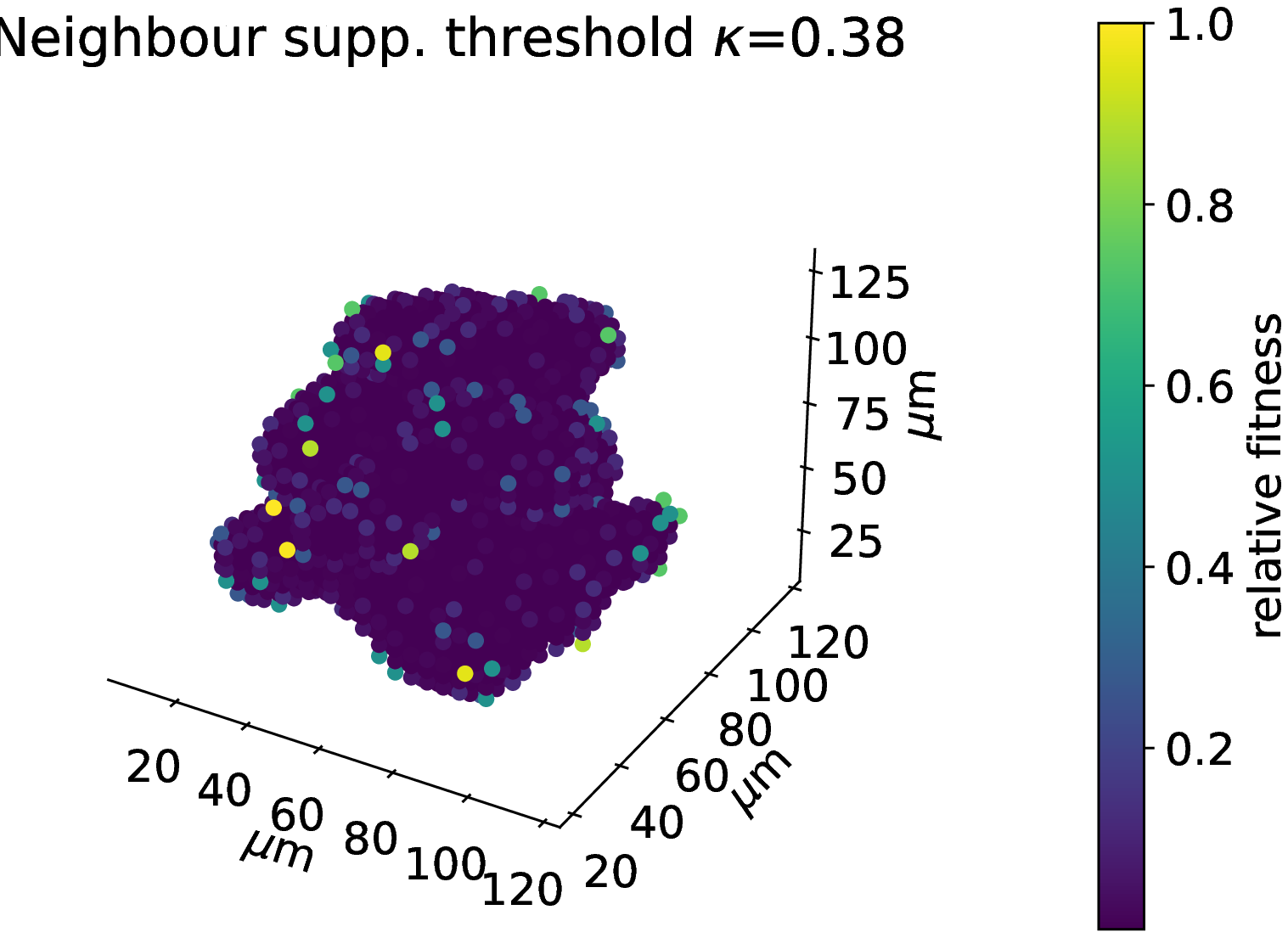}
    \includegraphics[width=.32\linewidth]{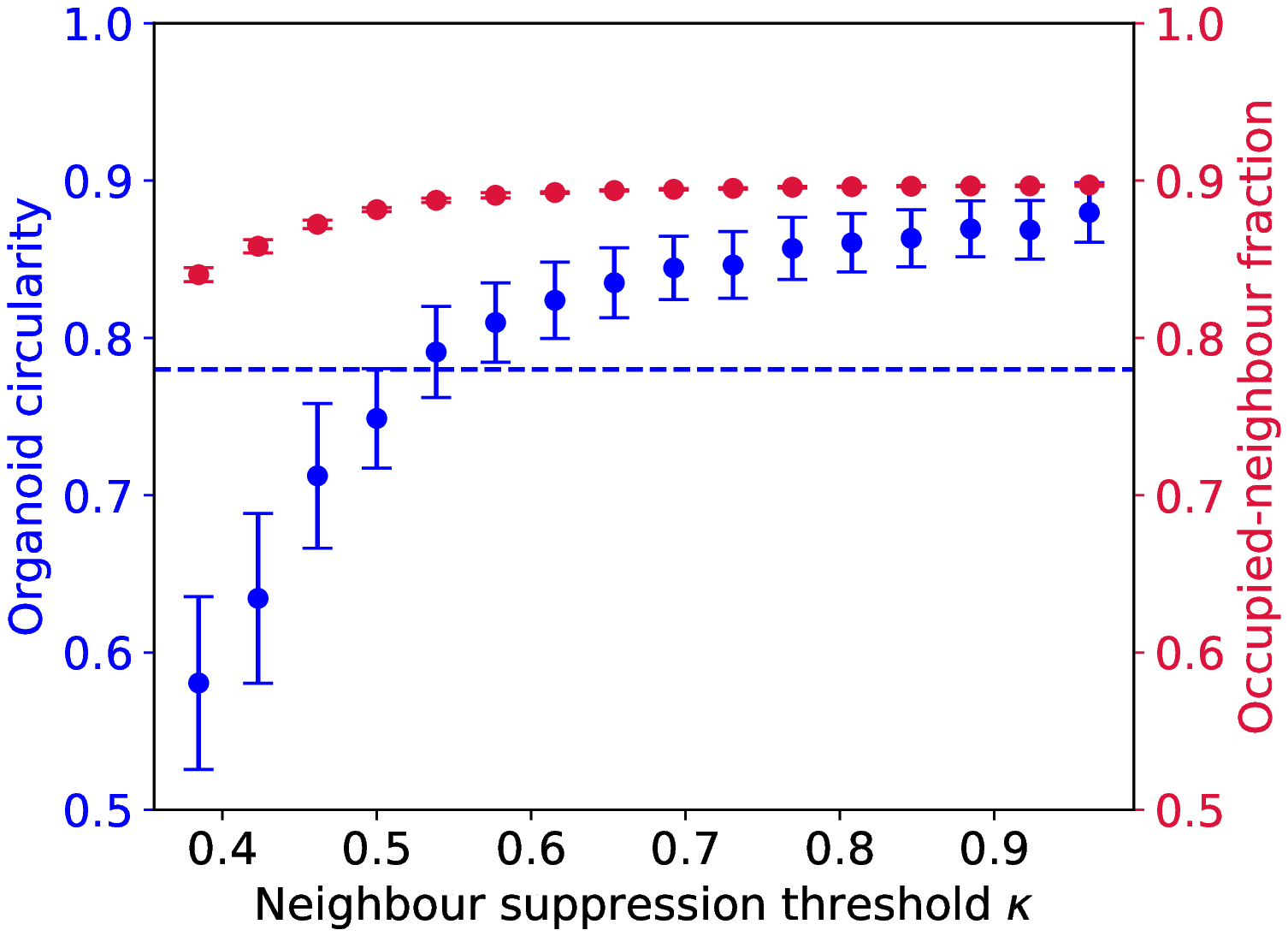}
    \includegraphics[width=.32\linewidth]{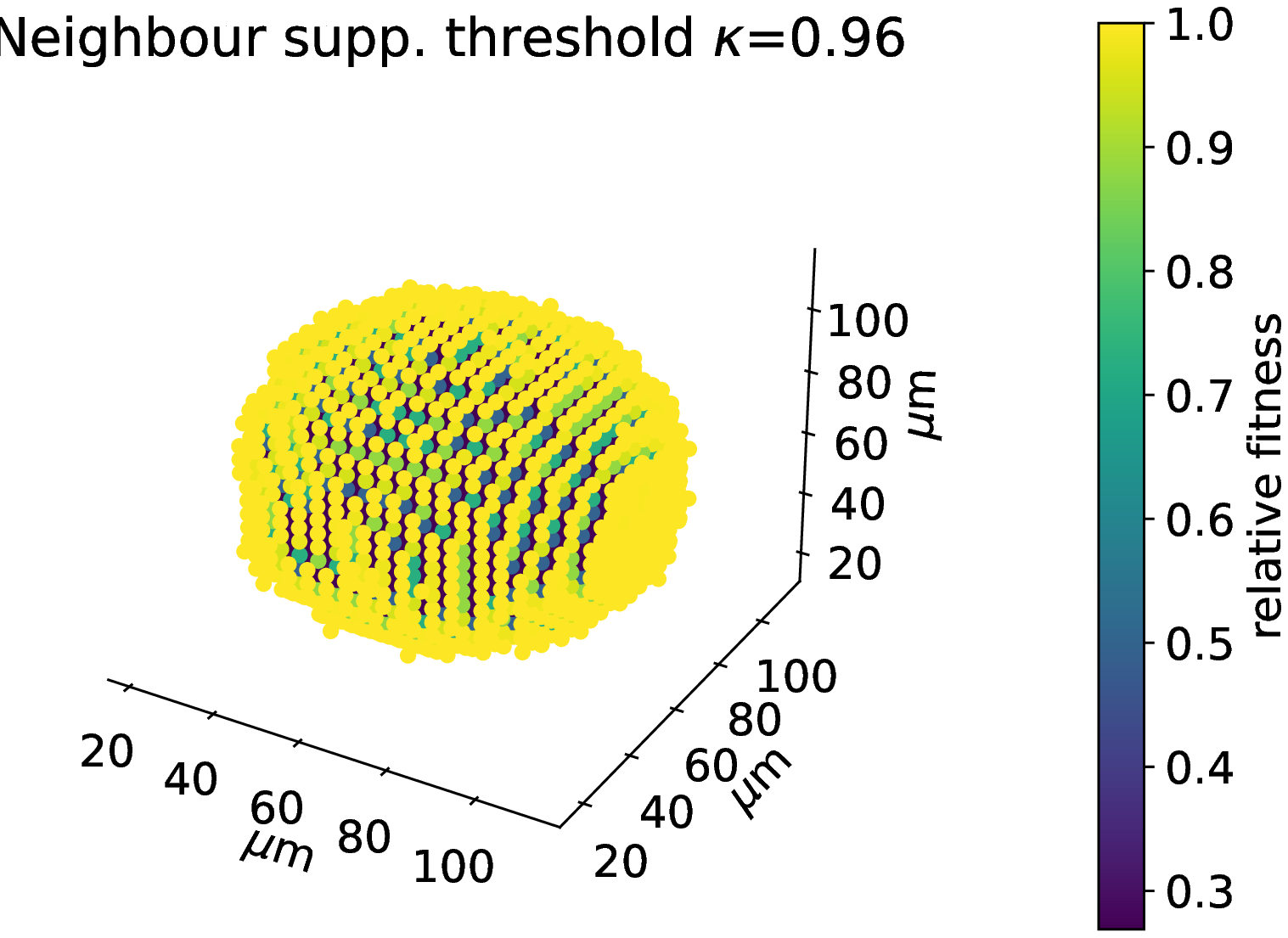}

  \caption{Illustrations of the effect of varying (first row) steepness \(h\), (second row) cell-cell adhesion \(\tau\), and (third row) threshold \(\kappa\). For all simulations \(\alpha=0, \beta=10\).  Default parameters are \(c=10, h=1, \tau=1\); all parameters are at these values unless stated otherwise above a graph. All axes are in \(\mu m\); here 5,000 cells gives an experimentally realistic diameter of \(100 \mu m\). Colour indications are as before.}
\end{figure}

We can also see that the baseline division rate of surrounded cells, \(\alpha\), must be negligible compared to \(\beta=10\) to allow the development of budding growth from Figure 7, below, which shows the emergence of mutant-like organoids as \(\alpha\) is decreased from \(0.1\) (near-spheroidal growth) to zero (budding growth). In order to stay beneath the mutant-like organoid threshold \(C=0.78\), the baseline division rate of the cells must be negligible (i.e. \(\alpha < 0.001\beta \)). 

\begin{figure}[h!]
    \includegraphics[width=\linewidth]{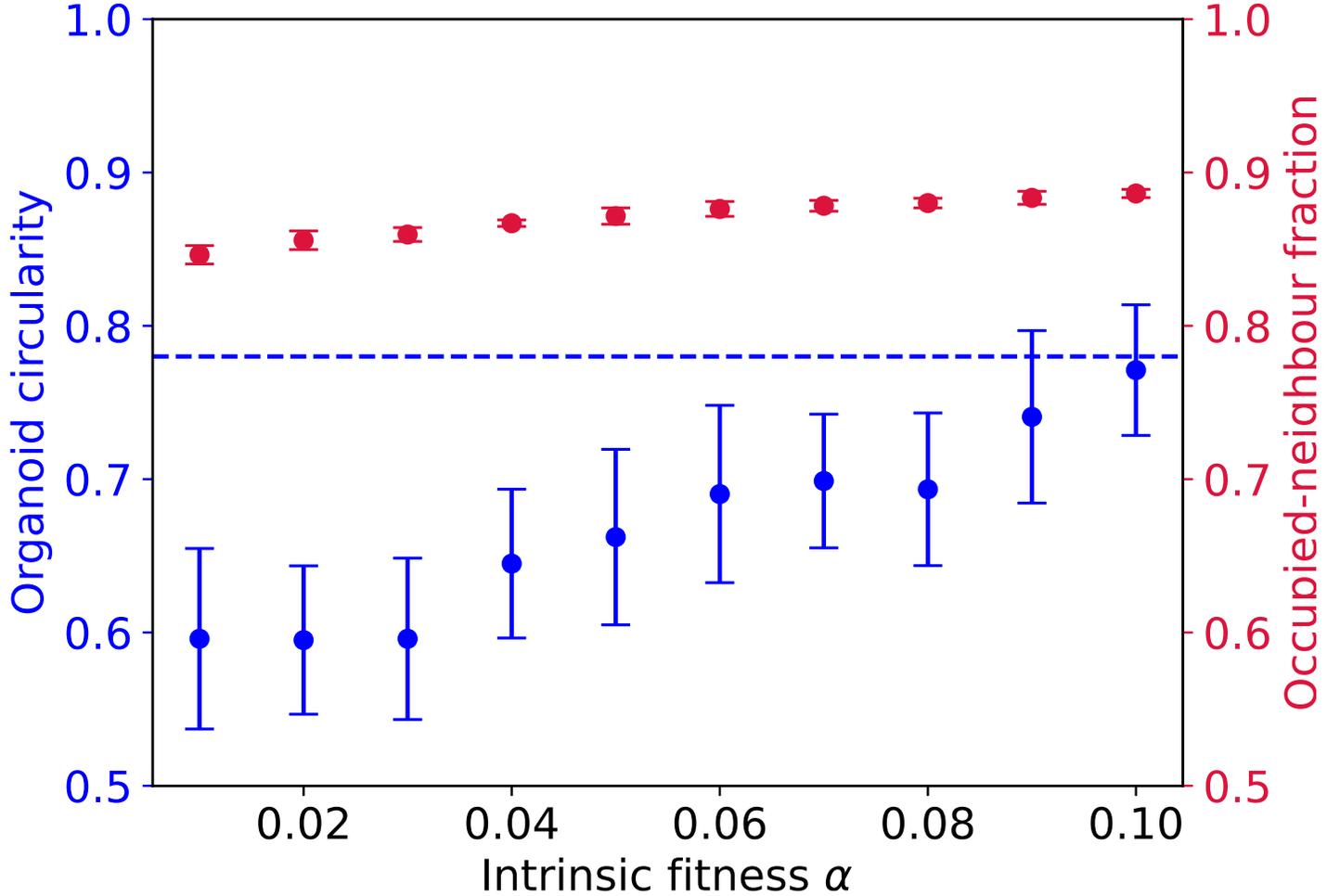}
  \caption{Illustration of the effect of varying inherent fitness \(\alpha\) in organoids grown with \(\beta = 10, \tau = 1.0, h = 1, c=10\).}
\end{figure}

   We make two further significant observations. Firstly, when neighbour suppression is high, the experimentally-witnessed diameter of about \(100 \mu m\) can be replicated with only 5,000 cells (versus roughly 10,000 without suppression, see previous section). Intuitively, this is the effect of protrusion-based growth: organoids with high neighbour suppression limit their division to surface cells and so grow outwards, as opposed to clustering together at the centre. In other words, neighbour suppression results in more invasive organoid behaviour. This could partly explain why carcinomas induced in mice with this mutation are so diffusive \cite{Politi2006}: neighbour suppression would force cells to grow away from each other, and a mutant cell at the boundary of an expanding clone would both be able to prevent the division of its wild-type neighbours and be sharply advantaged compared to a mutant cell at the organoid's centre.

    \subsubsection{Accumulation of suppression produces deformations but not mutant-like `budding'}
    The picture that emerges is now clearer. Our simulations suggest that cells in budding organoids are much fitter than cells in spheroidal organoids (represented by low \(\beta\) in our model), but surrounded cells are prevented from reproducing by contact with neighbouring cells (represented by a negligibly low baseline \(\alpha\)). Possible mechanisms for this suppression are discussed in the Conclusions. One important note is that the mechanism of suppression does not necessarily have to be exclusively current; cells which remember being surrounded can also produce deformations. To illustrate this, we can consider an alternative model, similar to the differentiation model in section 4.2.3, in which all cells are born active and become inactive in response to being surrounded. In this model all active cells have the same per-timestep division probability \(p_{\text{diff}}(n, dT) = \alpha dT\). However, in a given timestep, an active cell with \(n\) empty neighbouring lattice-points has a probability \(\frac{\sigma dT}{1+e^{h(c-(26-n))}}\) of becoming inactive, for some `inactivation rate' \(\sigma\). This probability is almost zero when the cell is isolated and maximal when it is surrounded. A cell which is surrounded for a long time may initially be active, but is very likely to eventually be forced into inactivity through prolonged exposure to other cells. Even if it is later pushed to the surface, it will remain inactive. We call this mechanism `suppression accumulation'. The results of simulations with similar `inhibition strength' parameters \(\alpha=10, c=10, h=1, \tau =1.0\) and various values of \(\sigma\) are shown in Figure 8. (We use a high value of \(\alpha\) here for computational efficiency, since continual inactivation will considerably increase the timescale of organoid growth without any corresponding increase in division rate.) For high rates of inactivation, we observe non-spherical deformations, enough to bring the circularity below the threshold \(C=0.78\) required to classify an organoid as `mutant-like' in some cases. However, visual inspection shows that these structures are in fact ellipsoidal, not budding. As with differentiation-based growth, the fact that active cells are only produced by other active cells means that proliferation can easily become lopsided; here small groups of active cells proliferate from an off-centre point within the cluster, resulting in an elongated `rugby-ball' shape. As cells beneath the surface are still able to reproduce, however, we do not see the development of protrusions. We conclude that, while cumulative suppression may exist within the structure, it is insufficient to produce truly `mutant-like' organoids.
\begin{figure}[H]
    \includegraphics[width=.32\linewidth]{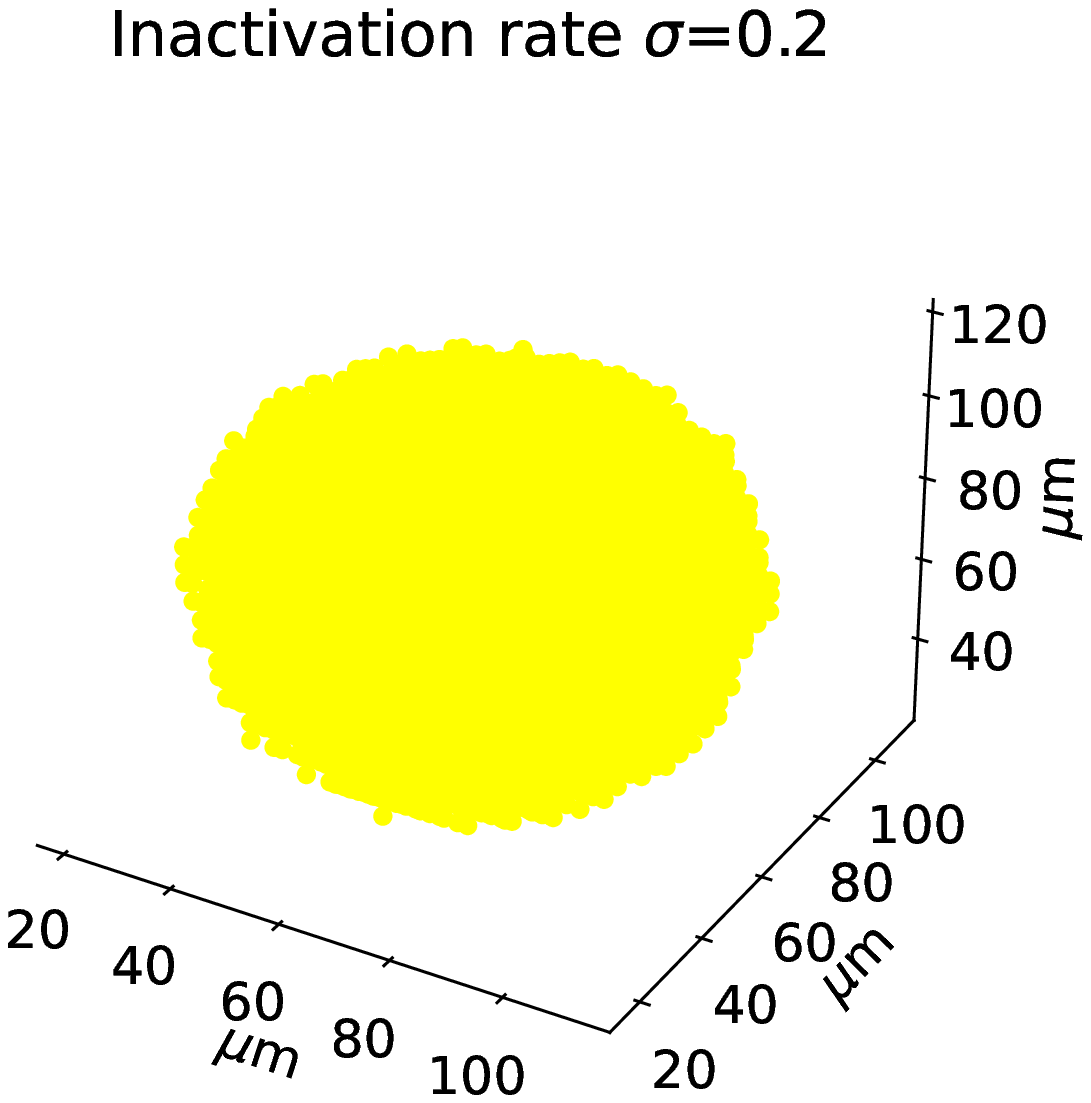}
  \includegraphics[width=.32\linewidth]{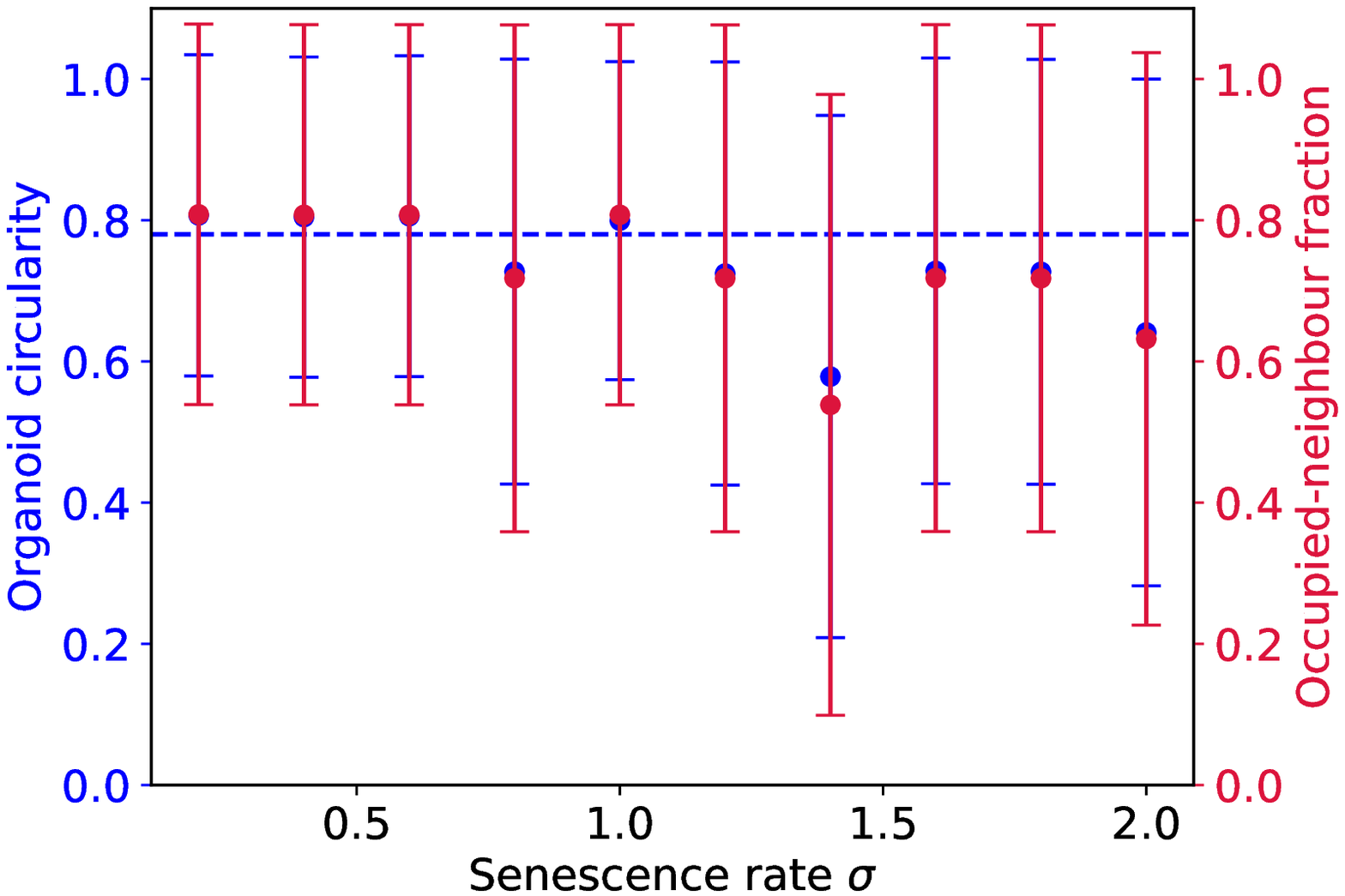}
  \includegraphics[width=.32\linewidth]{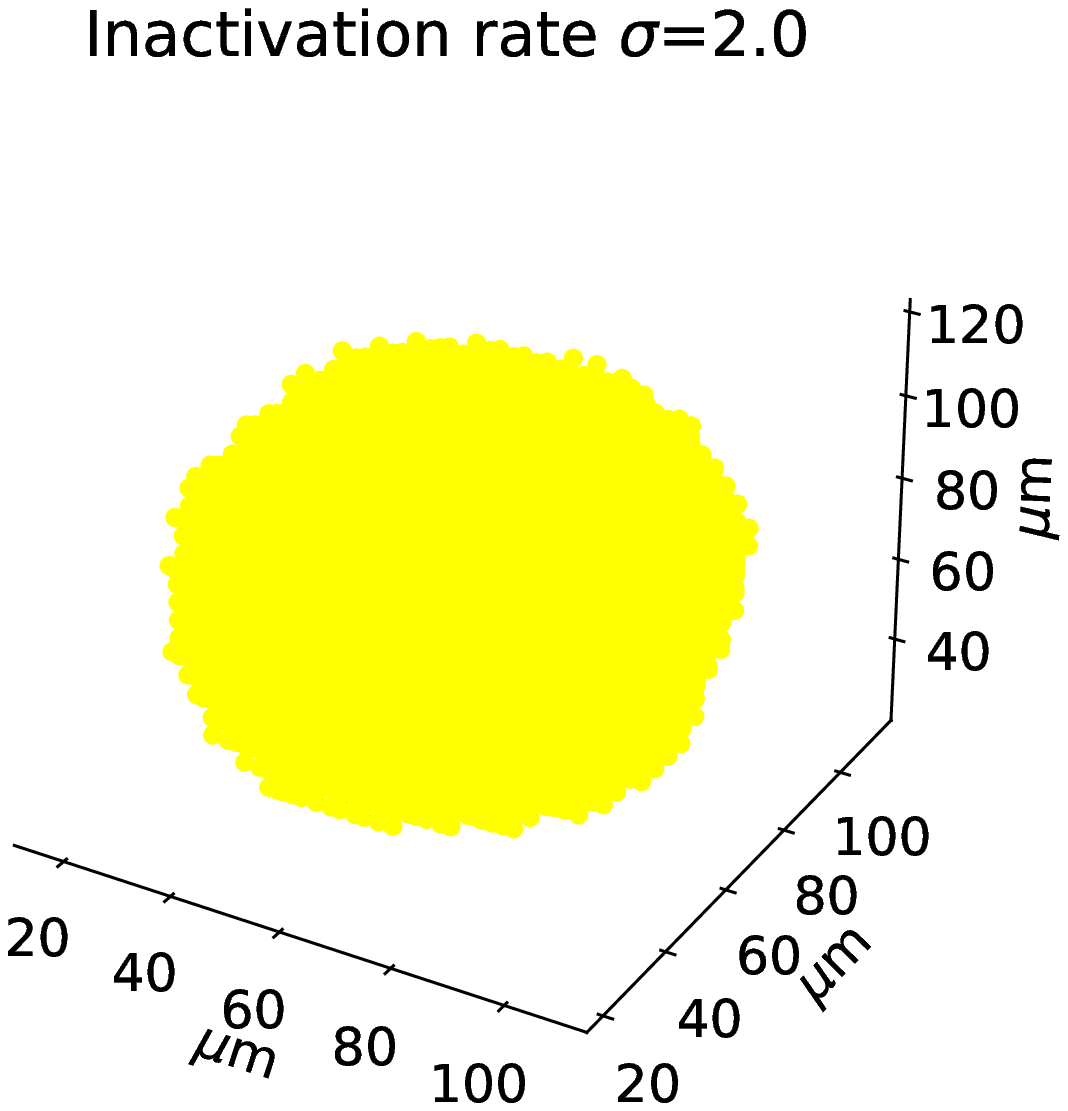}
  \caption{The effect of inactivation rate on `accumulated-inactivity'-driven growth as described above. Here \(\alpha=10,  h=\tau=1, c=10\) and \(\sigma\) is varied. All simulated organoids are run to 5000 cells. Here colour indicates the fitness a cell \textit{would} have were it active; the positions of active and inactive cells are not shown.}
  \end{figure}
  \subsection{Long-range neighbour suppression is sufficient to produce budding structures}

Finally, we relax the assumption that cells can only interact when their surfaces touch, and consider a model in which cells can interact at larger distances. This can happen because cells absorb nutrients within the Matrigel, depleting the resources available to another cell further away; or because cells actively secrete some inhibitory factor, whose concentration decreases with distance. Both hypotheses result in mathematically equivalent models. In the following explanation we will discuss this phenomenon in terms of active inhibition. We assume that cell \(i\) at position \(\underline x_{i}\) secretes some factor with concentration \(\rho_i(\underline x, t)\) that inhibits the growth of other cells (and that this may simply be `the absence of nutrients'). This factor is diffusive and so, outside the cell itself, obeys the diffusion equation (for some diffusion constant \(D\)):

\[\frac{\partial \rho_i}{\partial t} = D \nabla^2 \rho_i\]
We assume the nutrient, applied externally by the administrator, diffuses much more rapidly than it decays, and so we need not consider a degradation term (which would lead to an exponential solution in one dimension and a Bessel function form in two). We assume that on the scale of cell reproduction the concentration of this inhibitory factor has reached a steady state, and further that it is radially symmetric about the cell, so that at any point which is a distance \(r_i\) from the centre of cell \(i\) (assuming that \(r_i > r_0\), the cell radius) the inhibitory factor obeys the equation 

\[\nabla^2 \rho_i = \frac{1}{r_{i}^2}\frac{\partial}{\partial r_i}(r_{i}^2 \frac{\partial \rho_i}{\partial r_i}) = 0\]
We assume that \(\rho_i \rightarrow 0\) as \(r_i \rightarrow \infty\), i.e. that the influence of one cell on another vanishes at infinite separation. This leads to \(\rho_i = \frac{\lambda}{r_i}\) for some constant \(\lambda\), which we take to be the same for all cells within the same organoid. We further assume that cells act as point sources of this factor, that cells can be approximated by spheres and absorb this factor at their surfaces (where concentration is finite). We need only calculate inter-cellular influence, as the effect of each cell's secretions on its own division rate is absorbed into the baseline fitness. When calculating the concentration of inhibitory factor secreted by a distant cell at the surface of a focal cell, we neglect any variation in concentration across this surface and assume that the concentration is dependent only on the distance between the two cell centres. This leads to a fitness \(\alpha - \mu \Sigma_{j \neq i} \rho_j(\underline x_i)\) for cell \(i\), where \(\rho_j(\underline x_i)\) denotes the concentration of the inhibitory factor secreted by cell \(j\) at the position of cell \(i\). Adding this together, the probability that cell \(i\) will divide becomes

\[p_{\text{LR}}(dT) = \alpha(1- \eta \Sigma_{j \neq i} \frac{1}{r_{ij}}) dT \]
where \(r_{ij}\) indicates the distance between the centres of cell \(i\) and cell \(j\), and for some scaling factor \(\eta\) which we vary, and which we refer to as the `inhibition level'. We take \(\eta\) small enough that this is always positive in the two-cell case (i.e. to keep individual influences are small); within the simulation, this probability may drop below zero for a highly surrounded cell in a the case of strong depletion, in which case the cell simply cannot divide (i.e. has an effective \(p_{\text{LR}}=0\)). 

Figure 9 shows the effect of increasing this long-range inhibition. At \(\eta \approx 0.01\), long-range depletion is sufficient to create mutant-like organoids. As previously, this budding structure occurs because all but a small number of cells located in surface protrusions are prevented from dividing, even those in contact with the Matrigel. Figure 10 shows the dependence of this structure on organoid size (though this cannot be assessed quantitatively, as our image processing pipeline is unable to accurately define a perimeter for small organoids). Thus, the distinction between mutant and non-mutant organoids can be explained either by universal long-range depletion and by an increase in division rate amongst mutant organoids, such that only mutants reach a number of cells required to prompt the emergence of budding structures; or by an increase in long-range suppression by mutant cells, either through active inhibitory signalling or enhanced absorption of key nutrients. This latter possibility does not preclude an increase in division rate, but does not require it. We are unable to determine the number of cells included in a non-spherical structure through cross-sectional areas, so it is unclear from our data whether mutant organoids necessarily contain more cells than non-mutants. Further experiments should weigh all organoids to clarify this question. 

Whilst higher inhibition should decrease the overall amount of division in the organoid, it may not necessarily decrease the observed diameter, as the development of surface protrusions packs cells very inefficiently and allows rapid expansion into the surrounding Matrigel. In general, we can conclude that high levels of long-range inhibition are sufficient to concentrate division within surface protrusions and result in bubbling growth.

\begin{figure}[H]
    \includegraphics[width=.32\linewidth]{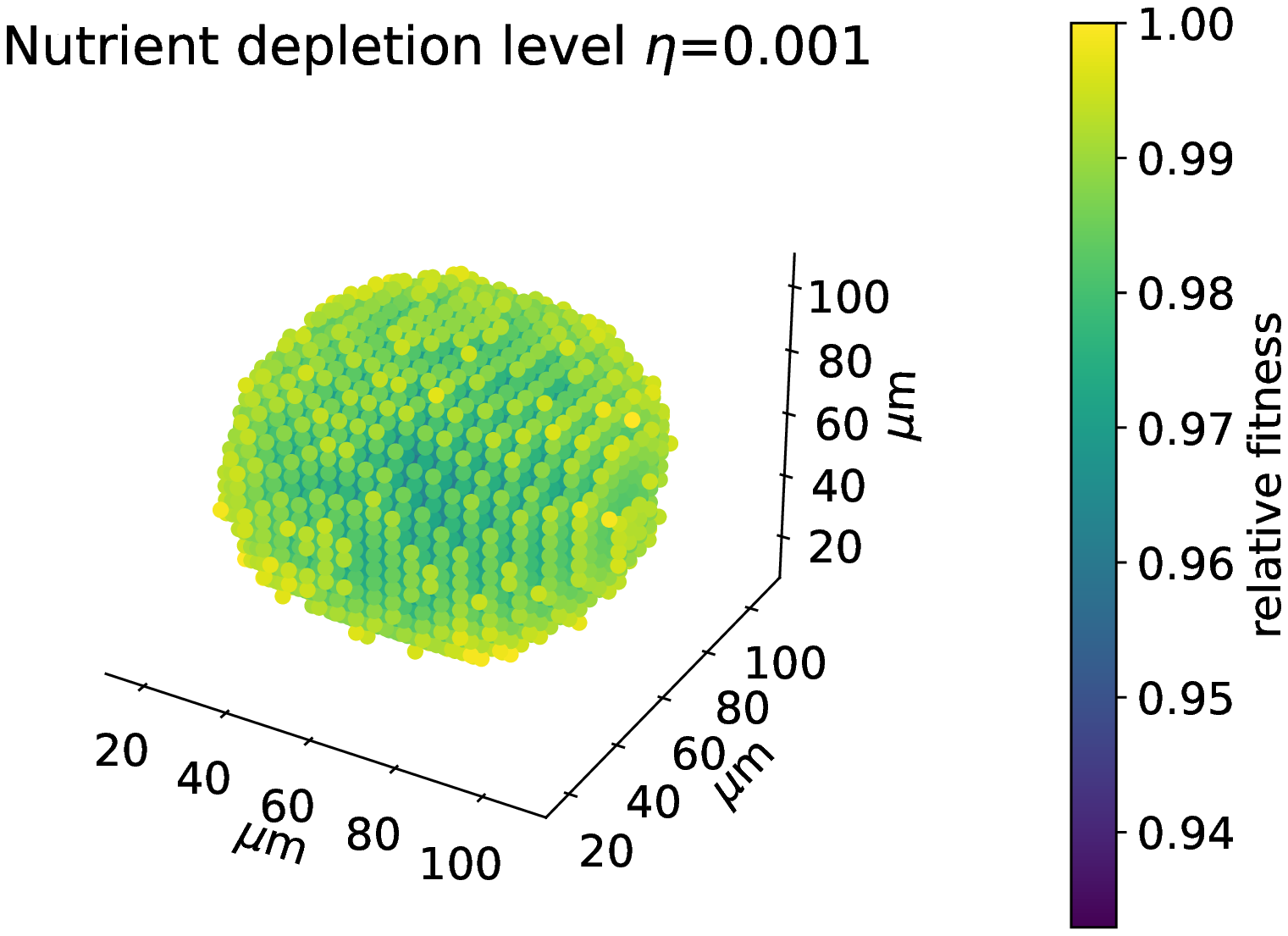}
  \includegraphics[width=.32\linewidth]{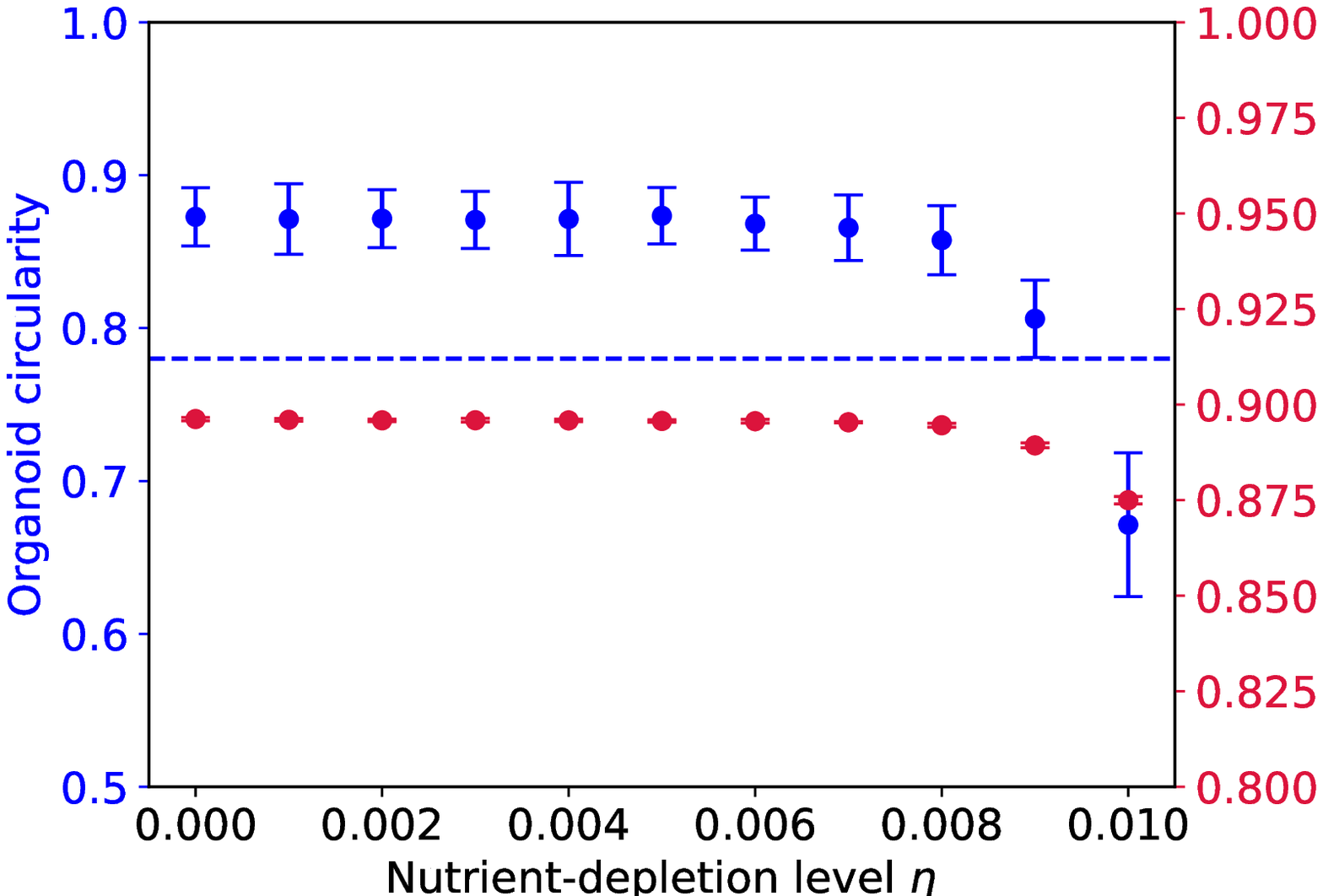}
  \includegraphics[width=.32\linewidth]{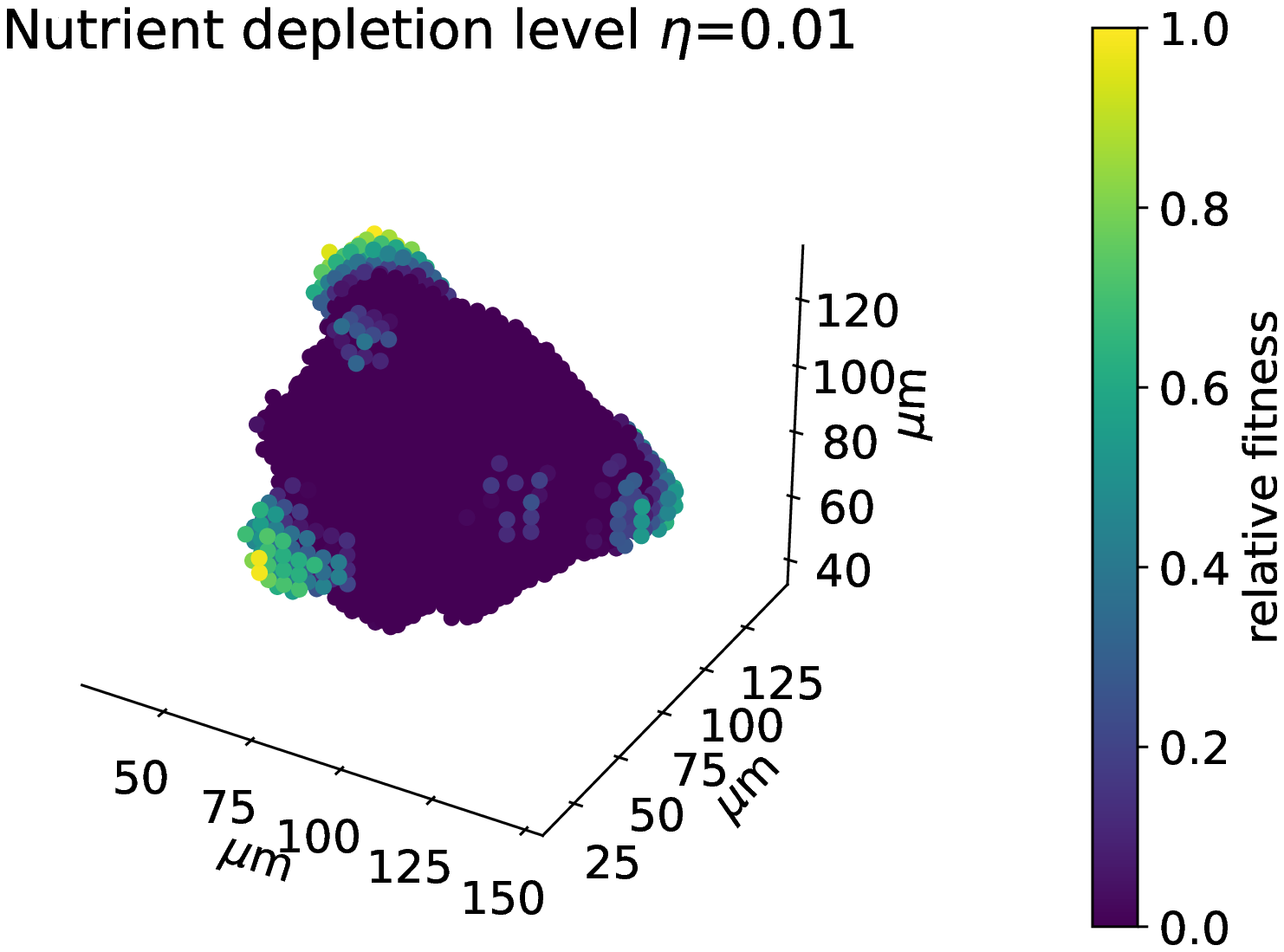}
  \caption{The effect of long-range inhibition on organoid growth. Here \(\alpha=11, \tau=1\) and \(\eta\) is varied. All simulated organoids are run to 5000 cells. Colour indicates the relative fitness of a cell.}
  \end{figure}

  \begin{figure}[H]
    \includegraphics[width=.32\linewidth]{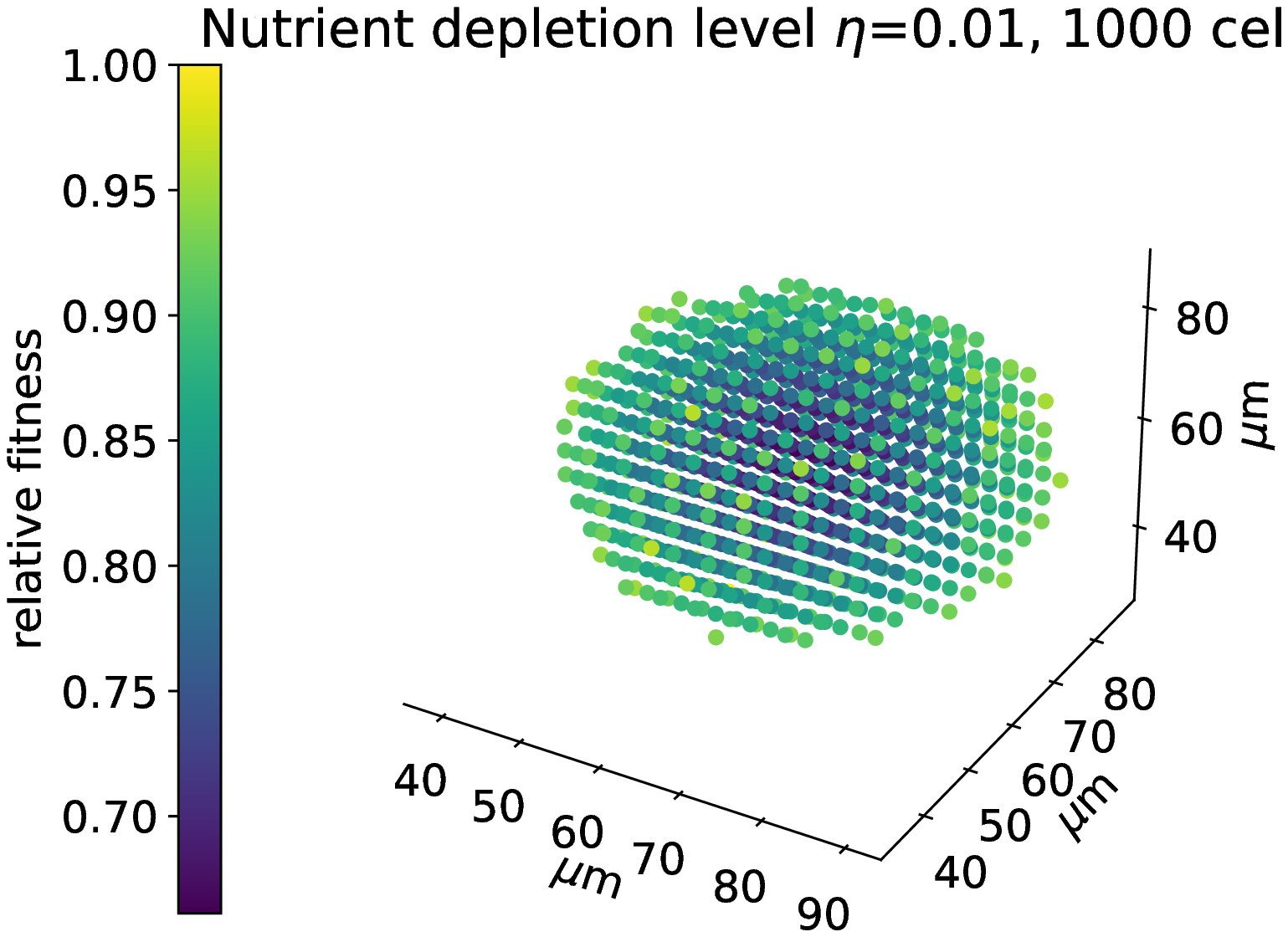}
  \includegraphics[width=.32\linewidth]{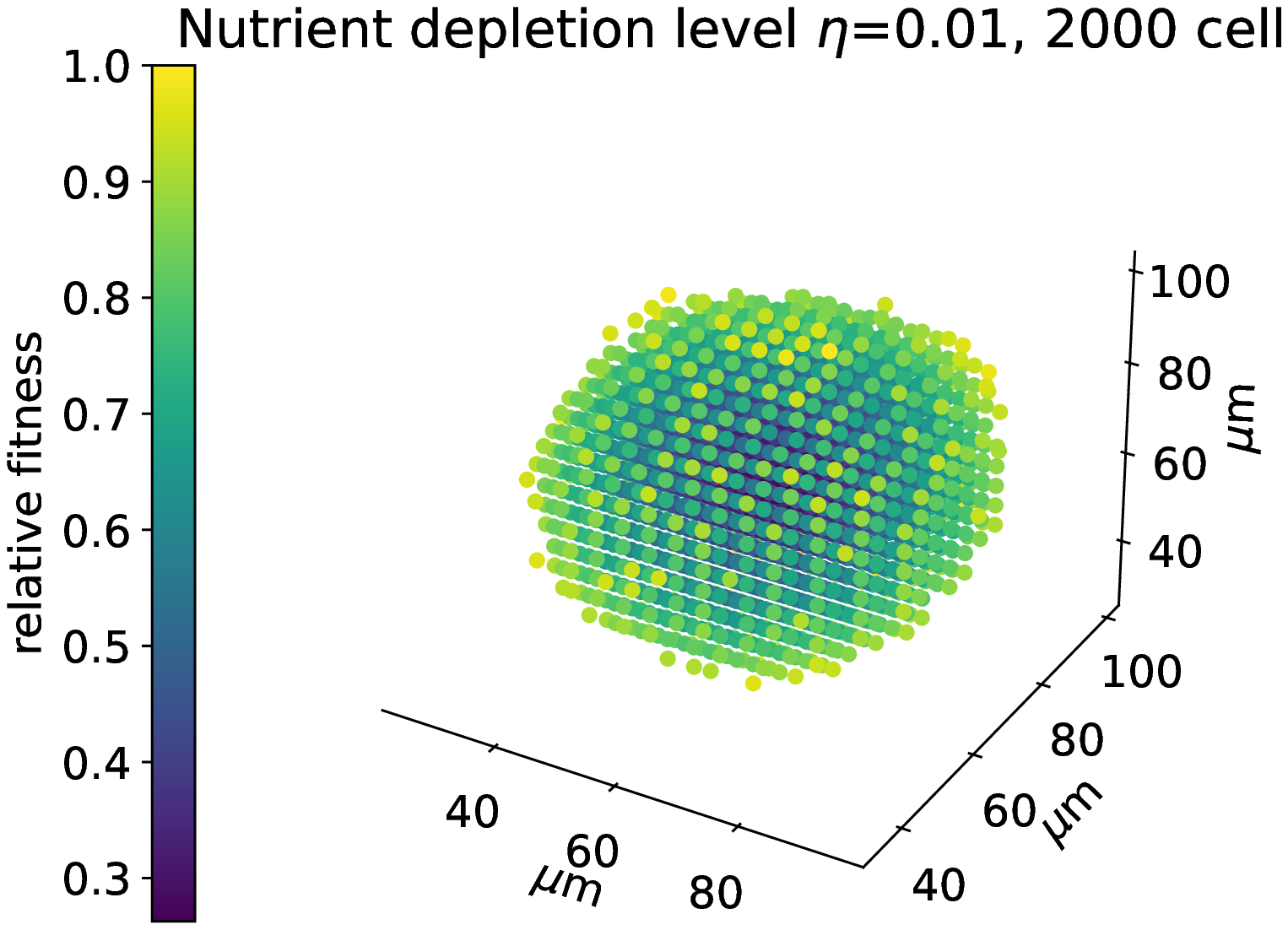}
  \includegraphics[width=.32\linewidth]{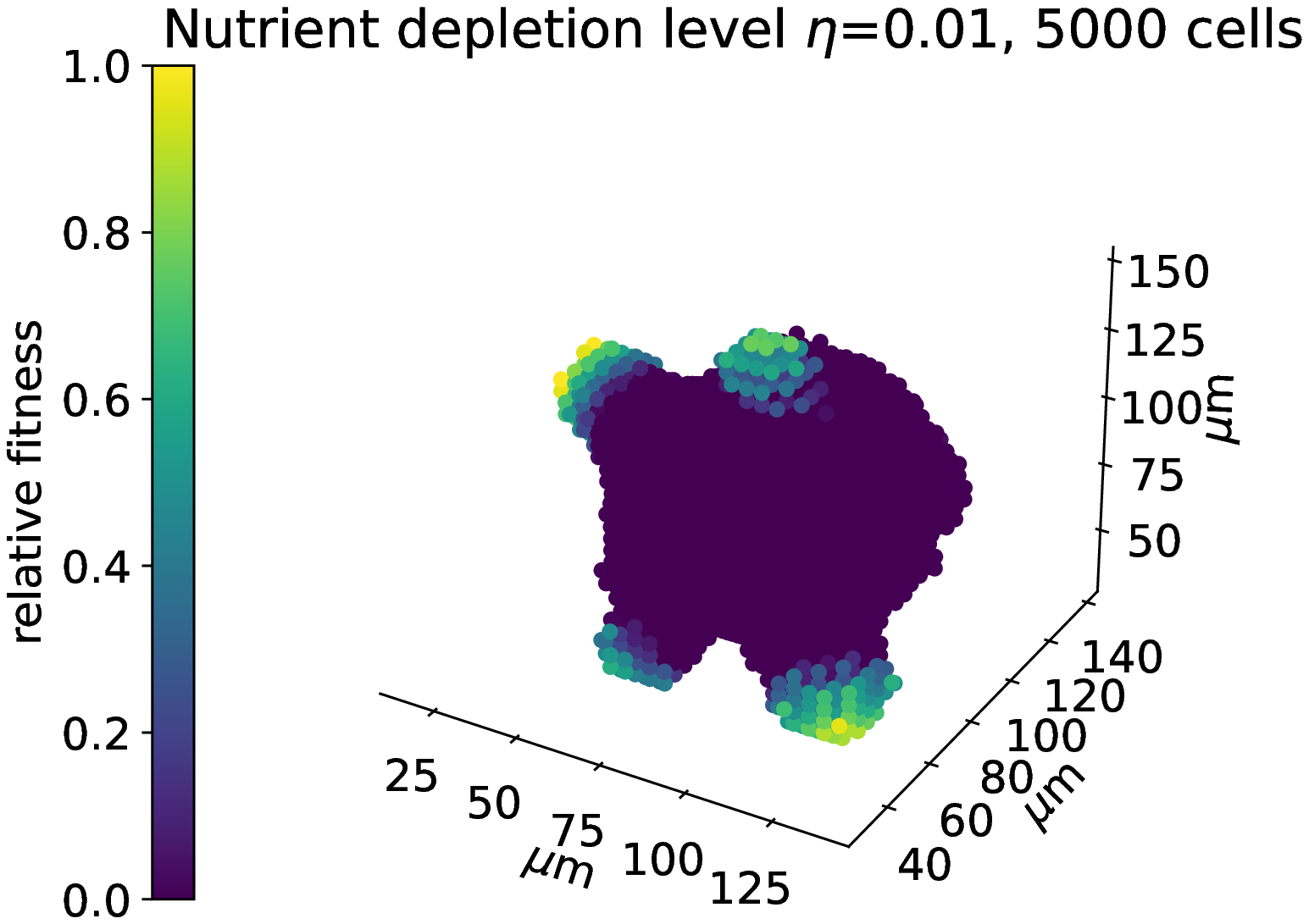}
  \caption{The effect of long-range inhibition on organoid growth. Here \(\alpha=11, \tau=1\) and \(\eta\) is varied. All simulated organoids are run to 5000 cells. Colour indicates the relative fitness of a cell.}
  \end{figure}

\section{Discussion and conclusion}

In this study we have developed an agent-based modelling approach to simulate the growth of mutant and non-mutant organoids. We have found that the secondary spheroids formed by mutant clusters can be explained by several hypotheses, all of which result in the restriction of division to a small number of surface cells in order to facilitate protrusion. This study builds upon the body of theoretical and computational work establishing the existence of instabilities in organoid growth \cite{Ciarletta2013, Miura2008, Giverso2016}, and our agent-based modelling approach allows us to straightforwardly calculate the detailed morphology of the resulting developed protrusions. 

We find that four possible models can predict the emergence of mutant-like organoids. In one (Hypothesis A), high levels of differentiation exist in all organoids, but mutant cells make more efficient use of available resources. In another (Hypothesis B), long-range inhibition of cell division allows budding structures to develop in mutant organoids but not non-mutants because of an increase in inherent division rate. In the third and fourth, the mutation induces either short (Hypothesis C) or long-range (Hypothesis D) inhibition of division between neighbouring cells. 

Whilst we are unable to distinguish between these mechanisms at present, we can suggest the following set of experiments which could be used to do so. To assess Hypothesis A, flow cytometry can be used to determine the levels of differentiation across all organoids, and the relationship between nutrient levels and division rate should be measured in both mutant and non-mutant cells. This hypothesis predicts that all cells should be highly differentiated and that the proliferation of mutant cells should be much more sensitive than non-mutants to differences in nutrient concentration. 

Hypotheses B-D all predict that cells can inhibit each other's division but are agnostic as to whether this is active (through the secretion of some inhibitor) or passive (through nutrient depletion). Hypothesis B predicts that mutant and non-mutant cells should be equally sensitive to nutrient depletion (and that mutant cells should have a generally higher division rate). This dependence might be strong (which suggests passive inhibition) or negligible (which suggests that inhibition is active, or at least independent of any nutrients supplied by the experimenter). Hypotheses C and D suggest that inter-cell influences are \textit{induced} by the mutation, which could be detected by co-culturing mutant and non-mutant cells and measuring their division rate with time-lapse video. If this inhibition is short-range (Hypothesis C), then division should be inhibited only for those cells directly in contact with mutant cells. If it is long-range (Hypothesis D), division will be suppressed for all non-mutant cells co-cultured with mutant cells in comparison with those cultured alone. In general we will be able to detect active inhibition if the division rate of a cell can be lowered by contact or co-culture with other cells, but not by an experimenter-induced decrease in nutrient levels.

The mechanism of this active inhibition, if present, is currently unclear. Contact-inhibited proliferation (CIP), the phenomenon whereby cells stop dividing in areas of high local density \cite{Levine1965}, is known to be mediated by EGF concentration \cite{Kim2009} and so might in theory be affected by the presence of an EGFR mutation. Variations in CIP were found to produce morphologically distinct organoid structures in the simulations of Karolak \textit{et. al.} \cite{Karolak2019}. However, CIP is generally lost during the development of cancer \cite{Hanahan2000}, and so it would be surprising if such a cancer-associated mutation as \textit{EGFR-L858R} were found to induce it. Instead, we expect either that mutants compete more for nutrients than non-mutants, or that mutant cells secrete an antagonist that halts the division of their immediate neighbours, wild-type or mutant. A similar phenomenon has recently been observed amongst intestinal stem cells with mutated \textit{Apc} by Flanagan and colleagues \cite{Flanagan2021}. If this is happening in our system, then this should give \textit{EGFR-L858R} mutant AT2 cells a significant competitive advantage over non-mutant cells \textit{in vivo} and when grown in co-culture, and at least partially explain the invasiveness of L858R-mutant carcinomas. Further experimental work is needed to verify this. If a signalling pathway which leads to neighbour suppression could be identified, this might be pharmacologically targetable and have implications for the prevention and treatment of NSCLC amongst never-smokers.

The existence of even passive inter-cell inhibition, if confirmed, would suggest that spatial environment strongly influences whether or not a pre-cancerous cell is able to develop into a lesion. Cell clusters in narrow confines will inhibit each other's division and stall the overall growth of the cancer. But given enough room, pre-cancerous cells can escape each other's influence, allowing the development of invasive protrusions. Our work emphasises the importance of considering spatial structure and cell-cell interaction in tumorigenesis, and to consider the circumstances of cell growth when determining whether or not a particular mutation will lead to cancer. Only by considering mutant cells within their full context, ecological and otherwise, will we be able to untangle the mechanisms that lead to cancer.

\section{Materials and Methods}

\subsection{Animal Procedures}

Animals were housed in ventilated cages with access to food and water ad libitum. All animal procedures were approved by The Francis Crick Institute Biological Research Facility Strategic Oversight Committee, incorporating the Animal Welfare and Ethical Review Body, conforming with UK Home Office guidelines and regulations under the Animals (Scientific Procedures) Act 1986 including Amendment Regulations 2012. Both male and female animals aged 6-15 weeks were used.

\textit{EGFR}-L858R [Tg(tet-O-EGFR-L858R)56Hev] mice were obtained from the National Cancer Institute Mouse Repository. Rosa26tTA and Rosa26-LSL-tdTomato mice were obtained from Jackson laboratory and backcrossed as previously described \cite{Hill2023}, \cite{Politi2006}. After weaning, the mice were genotyped (Transnetyx, Memphis, USA), and placed in groups of one to five animals in individually ventilated cages with a 12-hour daylight cycle. 

\subsection{Fluorescence-activated cell sorting} 
For flow cytometry sorting of alveolar type II cells, minced lung tissue was digested with Liberase TM and TH (Roche Diagnostics) and DNase I (Merck Sigma-Aldrich) in HBSS (Gibco) for 30 minutes at 37 $^{\circ}$ C in a shaker at 180 rpm. Samples were passed through a 100 \(\mu\) m filter, centrifuged (300 x g, 5 min, 4 degrees) and bed blood cells were lysed for 5 min on ice using ACK buffer (Life Technologies). Cells were blocked with anti-CD16/32 antibody (BD) for 10 min. Extracellular antibody staining was then performed for 30 min on ice (see table), followed by incubation in DAPI (Sigma Aldrich) to label dead cells. Cell sorting was performed on Influx, Aria Fusion or Aria III machines (BD). AT2 cells were defined as DAPI-CD45-CD31-Ter119-EpCAM+MHC Class II+ CD49f- as previously described \cite{Major2020}.

\begin{table}[]
\centering
\begin{tabular}{|l|l|l|l|l|}
\hline
\textbf{antigen} & \textbf{fluorochrome} & \textbf{Vendor} & \textbf{Cat \#} & \textbf{dilution} \\ \hline
CD45             & BV421                 & Biolegend       & 103134          & 1/150             \\ \hline
CD31             & BV421                 & Biolegend       & 102423          & 1/150             \\ \hline
Ter119           & BV421                 & Biolegend       & 1162          & 1/150             \\ \hline
EpCAM            & APC-Fire750           & Biolegend       & 118230          & 1/150             \\ \hline
MHC Class II     & FITC                  & Biolegend       & 107606          & 1/150             \\ \hline
CD49f            & PE-Cy7                & eBiosciences    & 25-0495-82      & 1/150             \\ \hline
\end{tabular}
\end{table}
\subsection{Organoid forming assay}

AT2 cells were isolated from control T or ET mice, without \textit{in vivo} Cre induction, incubated in vitro with 6 x 107 PFU/ml of Ad5-CMV-Cre in 100 \(\mu\)L per 100,000 cells 3D organoid media (DMEM/F12 with 10 percent FBS, 100 U ml-1 penicillin-streptomycin, insulin/transferrin/selenium, L-glutamine (all GIBCO) and 1mM HEPES (in-house)) for 1hr at 37$^{\circ}$ C as detailed in Dost \textit{et. al.} \cite{Dost2020}. Cells were washed three times in PBS, before 10,000 cells were mixed with a murine lung fibroblast cell line (MLg2908, ATCC, 1:5 ratio) and resuspended in growth factor reduced Matrigel (Corning) at a ratio of 1:1. 100 \(\mu\) l of this mixture was pipetted into a 24-well transwell insert with a 0.4 \(\mu\)m pore (Corning). After incubating for 30 min at 37 $^{\circ}$ C, 500 \(\mu\) l of organoid media was added to the lower chamber and media changed every other day, following previous methods \cite{Choi2020}. Bright-field and fluorescent images were acquired after 14 days using an EVOS microscope (Thermo Fisher Scientific) and quantified using FiJi (.2.0.0-rc-69/1.52r, ImageJ). For wholemount staining of organoids, organoids were prepared according to previous methods \cite{Dekkers2019} and stained with anti-proSPC (Abcam, clone EPR19839) and anti-keratin 8 (DSHB Iowa, clone TROMA-1). 3D confocal images were acquired using an Olympus FV3000 and analysed in FiJI.

\textbf{Computational analysis}

All organoids captured at sufficient resolution to accurately determine an area and perimeter were analysed using the scikit-image and Shapely libraries. Images were converted to grayscale and passed through a Gaussian filter with \(\sigma=0.8\), for all except the non-mutant organoid labelled `t2', which was captured at lower resolution and required \(\sigma=1.0\) for the perimeter to be found correctly. Contours were found using the function shapely.measure.find-contours with parameter `level' = 0.5; these were then visually verified. The concave hull of these contours was then computed using shapely.concave-hull with parameter `ratio' = 0.05. Circularity was determined using the variables hull.area and hull.length (to describe the perimeter). The same default parameters were applied to the analysis of all simulated organoids. When calculating the circularity of these organoids, 9 different images were taken from different viewpoint angles, using all possible combinations of elevation and azimuthal angles = \(-60 ^{\circ}, -20 ^{\circ}, 20 ^{\circ}\).

\section{Acknowledgements}

H. C. is supported by a grant from the Engineering and Physical Sciences Research Council [EP/W523835/1]. P. P. is supported by a UKRI Future Leaders Fellowship [MR/V022385/1]. M. P. D. is supported by the UK Engineering and Physical Sciences Research Council [EP/W032317/1]. C.S. acknowledges grant support from AstraZeneca, Boehringer-Ingelheim, Bristol Myers Squibb, Pfizer, Roche-Ventana, Invitae (previously Archer Dx Inc - collaboration in minimal residual disease sequencing technologies), and Ono Pharmaceutical. He is an AstraZeneca Advisory Board member and Chief Investigator for the AZ MeRmaiD 1 and 2 clinical trials and is also Co-Chief Investigator of the NHS Galleri trial funded by GRAIL and a paid member of GRAIL’s Scientific Advisory Board. He receives consultant fees from Achilles Therapeutics (also SAB member), Bicycle Therapeutics (also a SAB member), Genentech, Medicxi, Roche Innovation Centre – Shanghai, Metabomed (until July 2022), and the Sarah Cannon Research Institute C.S has received honoraria from Amgen, AstraZeneca, Pfizer, Novartis, GlaxoSmithKline, MSD, Bristol Myers Squibb, Illumina, and Roche-Ventana. C.S. had stock options in Apogen Biotechnologies and GRAIL until June 2021, and currently has stock options in Epic Bioscience, Bicycle Therapeutics, and has stock options and is co-founder of Achilles Therapeutics. K.M.P. is supported by Cancer Research UK (CRUK; CGCATF-2021/100014), the National Cancer Institute (CA278730-01) and the Mark Foundation via a Cancer Grand Challenges partnership (NexTGen).
Kate Gowers, Matthew A. Clarke, Yuxin Sun, Daniel Jacobson, Francesco Moscato, Tom Cox, Charlie Barker, Pedro Victori Rosa, Jasmin Fisher, Hugh Selway-Clarke, Sasha Bailey and Erik Sahai are thanked for their advice.

\bibliography{main.bib}
\bibliographystyle{vancouver}

\section{Data availability statement}
All data and code used in this work is provided in the Github repository \url{https://github.com/hcoggan/organoid_growth}. The EGFR-condition images have been previously published in \cite{Hill2023}; all other data is original.

\end{document}